\documentclass[12pt]{article}
\usepackage{graphicx,epsfig}

\usepackage{amssymb}
\usepackage{amsmath}

\newcommand{\bea}{\begin{eqnarray}}
\newcommand{\beq}{\begin{equation}}
\newcommand{\eea}{\end{eqnarray}}
\newcommand{\eeq}{\end{equation}}
\newcommand{\nn}{\nonumber}
\newcommand{\barr}{\begin{array}}
\newcommand{\earr}{\end{array}}
\newcommand{\Frac}[2]{\frac{\displaystyle{#1}}{\displaystyle{#2}}}
\newcommand{\lsim}{\raise0.3ex\hbox{$\;<$\kern-0.75em\raise-1.1ex\hbox{$\sim\;$}}}
\newcommand{\gsim}{\raise0.3ex\hbox{$\;>$\kern-0.75em\raise-1.1ex\hbox{$\sim\;$}}}

\newcommand{\eq}[1]{Eq.~(\ref{#1})}
\newcommand{\rhobar}{\bar {\rho}}
\newcommand{\etabar}{\bar{\eta}}

\begin{document}
\titlepage                                                          
\begin{flushright}                                                     
OUTP-0402P\\                                                       
MCTP-03-61\\
FTUV-04-0110
\end{flushright} 
\begin{center}                                                        
{\Large \bf Spontaneous $CP$ violation and Non-Abelian Family Symmetry 
in SUSY}\\
\vspace*{0.5cm}                                                          
Graham\ G.\ Ross{\footnote{g.ross1$@$physics.ox.ac.uk}}
\\{\small
Dep. of Physics, U. of Oxford, 1  Keble Road, Oxford,  OX1 3NP, UK and }\\
{\small Theory Division, CERN, CH-1211, Geneva  23, Switzerland}\\[.2cm]
Liliana Velasco-Sevilla{\footnote{lvelsev$@$umich.edu}}
\\{\small Michigan Center for Theoretical Physics, Randall Laboratory,
University of Michigan,}\\
{\small 500 E University Ave., Ann Arbor, MI 48109, USA 
and} \\
{\small  Dep. of Physics, U. of
Oxford, 1 Keble Road, Oxford, OX1 3NP, UK.}\\[.2cm]
Oscar Vives{\footnote{oscar.vives$@$cern.ch}}
\\ {\small Dep. de Fisica Teorica, U. de Valencia, Burjassot, E-46100, 
Spain and}\\
{\small Dep. of Physics, U. of Oxford, 1 Keble Road, 
Oxford, OX1 3NP, UK.}
\end{center}

\begin{abstract}

 We analyse the properties of generic models based on an $SU(3)$ family symmetry
 providing a full description of quark charged lepton and neutrino masses and mixing
 angles. We show that a precise fit of the resulting fermion textures is
 consistent with CP being spontaneously broken in the flavour
 sector. The CP violating phases are determined by the scalar potential and we discuss how symmetries readily lead to a maximal phase controlling CP violation in the quark sector. In a specific model the
 CP violation to be expected in the neutrino sector is related to that in the quark sector and we determine this relation for two viable models. In addition to giving rise to the observed structure of quark and lepton masses this class of model solves both the
  CP and  flavour problems normally associated with supersymmetric models. The flavour structure of the soft supersymmetry breaking terms is controlled by the family symmetry and we analyse some of the related phenomenological implications.

\end{abstract}
\section{Introduction}
\label{sec:intro}
The gauge sector in the Standard Model (SM) is completely fixed  by
gauge invariance once we specify the particle content and its quantum
numbers. This sector has been thoroughly tested  with
great precision at the collider experiments and the SM
 has been confirmed as the correct gauge theory of
strong and electroweak interactions up to energies of order 100 GeV.  
In contrast, the flavour sector of the SM remains the big
unknown in high energy physics. In the SM flavour (Yukawa) couplings
are not determined by symmetry and so they are merely unrelated 
parameters to be determined by experiment. In our search for a more fundamental
theory we need to improve our understanding of the
flavour physics. In analogy with the gauge sector, one might expect
a spontaneously broken family symmetry determines
the different flavour parameters. This approach has already
been tried several times with reasonable
degrees of success \cite{graham} although so
far it is difficult to choose one of the various options on offer.
The appearance of $CP$ violation in the SM is equally 
mysterious. We do not know why there are complex
couplings in the SM giving rise to a violation of the $CP$
symmetry in neutral meson systems. In fact, these two problems are
deeply related in the SM as the Yukawa couplings are the only
source for both the flavour structures and the $CP$ violation phenomena, 
although this need not be so in extensions of the SM where 
new sources of $CP$ violation different from the Yukawa couplings are
present. 

In this paper, we address both problems together and build
a supersymmetric flavour theory which determines all the flavour structure
of the theory as well as the different phases breaking the $CP$ symmetry. 
This is particularly relevant in the supersymmetric context where the 
lack of understanding of flavour and $CP$ is especially severe. 
The minimal supersymmetric extension of the SM (MSSM) has many new scalar 
states which introduce the possibility of new sources of flavour changing 
neutral currents (FCNC) beyond the usual Yukawa matrices
giving rise the so-called ``SUSY flavour problem''.
Similarly there are new $CP$ violating phases that, unless small, give 
unacceptably large contributions  to $CP$ violating observables -
the ``SUSY $CP$ problem'' \cite{annrev}. The simplest solution to these 
problems 
consists of arbitrarily assuming that all the new flavour structures are
proportional to the identity and that all new phases vanish. However,
this approach cannot be justified without some 
underlying symmetry reason (so far unspecified), made more difficult to 
understand given that the Yukawa
couplings of the SM do not share these features.  

We will study both $CP$ violation and the origin of flavour structures 
in the framework of a supersymmetric $SU(3)$ family symmetry model. This 
kind of model has been shown
\cite{king,Ross:2002fb,king2} to provide a correct texture for the Yukawa  
matrices in agreement with an accurate phenomenological fit \cite{romanino}.
However, the improvement on the determination of CKM mixing angles and  
$CP$ asymmetries requires the presence of phases in the elements of
the Yukawa matrices. In our analysis we assume that $CP$ is an exact
symmetry \footnote{In string theory CP invariance can be a symmetry of 
the 4D compactified theory corresponding 
to an unbroken discrete subgroup of the higher dimensional Lorentz group 
\cite{dine}}  of the theory of unbroken flavour $SU(3)$ and is only
spontaneously broken by complex vacuum expectation values of the flavon fields 
that determine the Yukawa structures \cite{spontCPflavour}. As we will 
discuss this approach has the great advantage of naturally explaining the 
SUSY CP problem. Moreover the CP violating phases are determined at the 
stage of flavour symmetry breaking and we will discuss how maximal CP 
violation naturally results.

In Section \ref{sec:su3} we present the general features shared by
$SU(3)$ flavour models which reproduce the observed quark textures. In
Section \ref{sec:quark} we show that spontaneous $CP$ violation in the
flavour sector is able to describe quark masses and mixing angles
while generating the observed $CP$ violation. In Section
\ref{sec:neutrino} we show how the same structure can reproduce the
correct masses and mixing angles in the leptonic sector. We give two
different examples of $SU(3)$ models which generate acceptable
neutrino masses and mixings and determine the prediction for the
different $CP$ violation observables in the lepton sector. In Section
\ref{sec:SUSY} we show that these models solve both the supersymmetric
flavour problem and the supersymmetric $CP$ problem and we point out
other observables where signals of a supersymmetric $SU(3)$ flavour
model may show up in future experiments. Then, in section \ref{sec:vac}
we analyse the vacuum alignment giving rise to the quark and lepton 
flavour structures and finally we present our conclusions.

\section{$SU(3)$ flavour symmetries and Yukawa textures}

\label{sec:su3}

One reason why it is difficult to construct a theory of flavour is that
measurement in the quark sector of the eigenvalues (quark masses) and the
CKM mixing matrix is all the information we can extract about the full quark
Yukawa matrices using SM interactions alone. Unfortunately, this is not
enough to determine the full structure of these matrices. Therefore there
remains a lot of freedom corresponding to widely different Yukawa textures,
ranging from democratic to strongly hierarchical matrices which, in the SM,
give rise to identical phenomenology. However, in most extensions of the
Standard Model, and in particular in supersymmetric extensions, the new non
SM interactions can distinguish between the different options.

Before the discovery of new supersymmetric interactions or any other new
interactions able further to probe the flavour sector we must rely on
simplifying assumptions to try to disentangle the complicated structure of
masses and mixing angles. A particularly interesting approach is to look for
texture zeros \cite{RRR}, small entries in the Yukawa textures which lead to
viable relations among the eigenvalues and mixing angles. An example is the
well known Gatto-Sartori-Tonin relation \cite{gatto} and its
phenomenological success may indicate some underlying dynamics capable of
generating those zeros. Given the success of the GST relation we consider it
reasonable to look for textures with this property. Moreover the smallness
of left handed mixing angles suggests that the non-diagonal elements
 are smaller than the diagonal
elements. A recent phenomenological analysis \cite{romanino} under these
assumptions shows that the following symmetric textures give an excellent
fit to quark masses and mixing angles, 
\begin{eqnarray}  \label{fit}
Y_d\propto\left( 
\begin{array}{ccc}
0 & b\ \bar \varepsilon^{3} & c\ {\ \bar \varepsilon^{3}} \\ 
. & {\bar \varepsilon^{2}} & a \ {\ \bar \varepsilon^{2}} \\ 
. & . & 1%
\end{array}%
\right),~~~~~~~~~~ Y_u\propto \left( 
\begin{array}{ccc}
0 & b^\prime{\ \varepsilon^{3}} & c^\prime{\ \varepsilon^{3}} \\ 
. & {\ \varepsilon^{2}} & a^\prime \varepsilon^{2} \\ 
. & . & 1%
\end{array}
\right)
\end{eqnarray}
with $\bar \varepsilon\simeq 0.15$, $\varepsilon \simeq 0.05$, $b=1.5$, $%
a=1.3$, $c=0.4$ and $a^\prime,b^\prime,c^\prime$ are poorly fixed from
experimental data. Note that with the assumption of small off-diagonal elements the data determines those 
entries above the diagonal (in our convention $Q_i d^c_j Y^d_{ij}$). 

In this paper we adopt this basic structure as our starting point.
As shown in \cite{king,king2} these structures can be successfully
reproduced from a spontaneously broken $SU(3)$ family symmetry. Under this
symmetry all left handed fermions ($\psi_i$ and $\psi^c_i$) are triplets. To
allow for the spontaneous symmetry breaking of $SU(3)$ it is necessary to
add several new scalar fields which are either triplets ($\overline{\theta}%
_{3}$, $\overline{\theta}_{23}$, $\overline{\theta}_{2}$) or antitriplets ($%
\theta_{3}$, $\theta_{23}$). We assume that $SU(3)_{fl}$ is broken in two
steps. The first step occurs when $\theta_3$ gets a large vev breaking $SU(3)
$ to $SU(2)$. Subsequently a smaller vev of $\theta_{23}$ breaks the
remaining symmetry. After this breaking we obtain the effective Yukawa
couplings at low energies through the Froggatt-Nielsen mechanism \cite{FN}
integrating out heavy fields. In fact, the large third generation Yukawa
couplings require  a $\theta_3$ (and $\bar \theta_{3}$) vev of the order of the
mediator scale, $M_f$, 
while $\theta_{23}/M_f$ (and $\bar \theta_{23}/M_f$) have
vevs of order $\varepsilon$ in the up sector and $\bar \varepsilon$ in the
down sector with different mediator scales in both sectors. 
Through vacuum alignment the D-terms ensure that the fields 
$\theta_{23}$ and $\bar \theta_{23}$ get equal vevs in the second and third
components. This is necessary to generate the Yukawa structure of 
Eq.~(\ref{fit}). Following \cite{king2}, we assume 
that $\theta_{3}$ and $\overline{\theta}_{3}$ transform as $\mathbf{3} 
\oplus \mathbf{1}$ under $SU(2)_R$ necessary to differentiate the up and down
Yukawas. The $SU(2)_R$ arises naturally in the context of an underlying $%
SO(10)$ grand unified theory and with this it is possible simultaneously to
describe quark and lepton Yukawa matrices. The main ingredient to reach this
goal, is a new field (or composite of fields) $\Sigma$, which is a $\mathbf{%
\overline{45}}$ of $SU(5)$ with a vev along the $(B-L+\kappa T_3)$ 
direction with $\kappa=2$,
corresponding to weak hypercharge. This generates the usual Georgi-Jarlskog 
factors that correct the difference
between down quark and charged lepton Yukawas.

\subsection{The superpotential}

The $SU(3)$ and $SO(10)$ symmetries are not enough to reproduce the textures
in Eq.~(\ref{fit}) and we must impose some additional symmetries to forbid
unwanted terms in the effective superpotential. The choice of these
symmetries is not uniquely defined (two examples are presented in
Section 4). Nevertheless the quark Yukawa textures consistent with the
requirement of $SO(10)\times SU(3)$ unification strongly constrain the
superpotential describing the Dirac Yukawa couplings. In an $SO(10)\times
SU(3)$ model, third generation Yukawa couplings are generated by $\theta
_{3}^{i}\theta _{3}^{j}$ while couplings in the 2--3 block of the Yukawa
matrix are always given by $\theta _{23}^{i}\theta _{23}^{j}\Sigma $. We
have two options for the couplings in the first row and column of the Yukawa
matrix. They can be given by $\epsilon ^{ikl}\overline{\theta }_{23,k}%
\overline{\theta }_{3,l}\theta _{23}^{j}\left( \theta _{23}\overline{\theta
_{3}}\right) $ or $\epsilon ^{ikl}\overline{\theta }_{23,k}\overline{\theta }%
_{3,l}\theta _{23}^{j}\left( \theta _{3}\overline{\theta _{23}}\right) $,
although the requirement of correct Majorana textures favours the first
option in all the cases analysed, as shown in Section \ref{sec:neutrino}.
With this the basic structure of the Yukawa superpotential is given by 
\footnote{The coefficients of $O(1)$ of the various terms have been suppressed for clarity}
\begin{eqnarray}
W_{Y} &=&H\psi _{i}\psi _{j}^{c}\left[ \theta _{3}^{i}\theta
_{3}^{j}\,+\,\theta _{23}^{i}\theta _{23}^{j}\Sigma \,+\,\epsilon ^{ikl}%
\overline{\theta }_{23,k}\overline{\theta }_{3,l}\theta _{23}^{j}\left(
\theta _{23}\overline{\theta _{3}}\right) \,+\,\right.   \nonumber \\
&&\left. \epsilon ^{ijk}\overline{\theta }_{23,k}\left( \theta _{23}%
\overline{\theta _{3}}\right) ^{2}\,+\,\epsilon ^{ijk}\overline{\theta }%
_{3,k}\left( \theta _{23}\overline{\theta _{3}}\right) \left( \theta _{23}%
\overline{\theta _{23}}\right) \,+\,\dots \right]   \label{YukawaW0}
\end{eqnarray}%
This structure is quite general for the different $SU(3)$ models we can
build, although in particular examples we may have different subdominant
contributions that can play a relevant role in some observables. As the
structure of Eq.~(\ref{YukawaW0}) fixes the main features of mixings,
eigenvalues and $CP$ phases we can make some general statements about the
phenomenology of such models.

On minimisation of the scalar potential the $SU(3)$ symmetry aligns the vevs
giving the form 
\begin{eqnarray}
\theta_3=\left( 
\begin{array}{c}
0 \\ 
0 \\ 
1%
\end{array}
\right)\otimes \left(%
\begin{array}{cc}
a_3^u & 0 \\ 
0 & a_3^d~ e^{i \chi}%
\end{array}
\right); & \bar{\theta}_3=\left( 
\begin{array}{c}
0 \\ 
0 \\ 
1%
\end{array}
\right)\otimes \left(%
\begin{array}{cc}
a_3^u~ e^{i \alpha_u} & 0 \\ 
0 & a_3^d~ e^{i \alpha_d}%
\end{array}
\right);  \nonumber \\
{\theta}_{23}=\left( 
\begin{array}{c}
0 \\ 
b_{23} \\ 
b_{23}~e^{i\beta_3}%
\end{array}
\right);& \bar{\theta}_{23}=\left( 
\begin{array}{c}
0 \\ 
b_{23}~e^{i\beta^{\prime}_2} \\ 
b_{23}~e^{i(\beta^{\prime}_2 - \beta_3)}%
\end{array}
\right);
\end{eqnarray}
As we will see in section \ref{sec:vac}, these vevs are of a form such that, 
\begin{eqnarray}
\label{expansion}
\frac{\displaystyle{a_3^u}}{\displaystyle{M_u}} = \frac{\displaystyle{a_3^d}%
}{\displaystyle{M_d}} = \varepsilon^{1/4}, ~~~~~~~ \frac{\displaystyle{b_{23}%
}}{\displaystyle{M_u}} = \varepsilon, ~~~~~~~ \frac{\displaystyle{b_{23}}}{%
\displaystyle{M_d}} = \bar \varepsilon.
\end{eqnarray}
where, to fit the masses and mixing angles, $\bar \varepsilon \simeq 0.15$
and $\varepsilon \simeq 0.05$ at the symmetry breaking scale that we take
approximately equal to $M_{GUT}$. Notice that, although $a_3/M =
\varepsilon^{1/4}$ is quite large as an expansion parameter, due to the 
$SU(3)$ symmetry it only enters
in the low energy parameters as $(\theta_3/M)^2 = \varepsilon^{1/2}$ or 
$(\theta_{23} \bar \theta_{3})/M^2$ (see for instance \eq{YukawaW0}). 
Then, the most 
important parameter for the hierarchy of fermion masses is $\theta_{23}$
which is quite small and the series expansion converges rapidly. So, with these
vevs we have the following Yukawa textures, 
\begin{eqnarray}  \label{thematrix}
Y^{f}=\left( 
\begin{array}{ccc}
0 & \varepsilon^3e^{i\delta}\, X_1 & \varepsilon^3e^{i(\delta^f + \beta_3)}\,
X_2 \\ 
... & \varepsilon^2\, \frac{\displaystyle{\Sigma}}{\displaystyle{|a^f_3|^2}} & 
\varepsilon^2 e^{i\beta_3}\, \frac{\displaystyle{\Sigma}}{\displaystyle{%
|a^f_3|^2}} \\ 
... & ... & e^{2 i \chi}%
\end{array}
\right) \frac{|a^f_3|^2}{M_f^2} ,
\end{eqnarray}
where $X_1$ and $X_2$ account for the different contributions to the $(1,2)$
and $(1,3)$ elements from the last three terms in Eq.~(\ref{YukawaW0}) and $%
\delta^{u,d}= (2 \alpha_{u,d} + \beta_3 +\beta^{\prime}_2)$. As mentioned
above the vev of $\Sigma$ changes for different fermion species. The vev $a^f_3
$ and the mediator masses, $M_f$, are different for fermions in the up or down
sector while for the up quarks and neutrinos $\chi = 0$.

In the same way, the supergravity K\"ahler potential receives new
contributions after $SU(3)_{fl}$ breaking. As explained in the introduction,
we assume that $CP$ is an exact symmetry of unbroken $SU(3)_{fl}$ and it is
only spontaneously broken by the complex flavon vevs. Thus, at $M_{Planck}$
both $SU(3)_{fl}$ and $CP$ are exact symmetries of the theory and both the
superpotential and the K\"ahler potential are invariant under these
symmetries. Furthermore, we use a Giudice-Masiero mechanism to generate a $%
\mu$-term of the correct magnitude so we have in the K\"ahler potential a
real coupling mixing the two Higgs doublets. In this class of theory the
hidden sector F-term breaking SUSY is also real. Therefore, after SUSY
breaking, we obtain a real $\mu$-term. In Section \ref{sec:SUSY} we show
that the corrections after $SU(3)_{fl}$ breaking do not modify this
conclusion.

\subsection{The K\"{a}hler potential}

After the flavour symmetry is spontaneously broken we obtain the Yukawa
structure of Eq.~(\ref{thematrix}). In the same way we can calculate the
effective K\"{a}hler potential that will be a general nonrenormalizable real
function invariant under all the symmetries of the theory coupling the
fields $\psi _{i}^{\dagger }\psi _{j}$ and flavon fields. In first place we
must notice that a term $\psi _{i}^{\dagger }\psi _{i}$ is clearly invariant
under gauge, flavour and global symmetries and hence give rise to a family
universal contribution. However, $SU(3)$ breaking terms give rise to
important corrections \cite{softflavour,alignment}. Any combination of
flavon fields that we add must be also invariant under all the symmetries,
although the $SU(3)$ indices can also be contracted with quark and lepton
fields. In fact, it is interesting to notice that, due to the possibility of
introducing the hermitian conjugate fields in the K\"{a}hler potential, new
terms are allowed with different suppression factors from the terms that
appear in the Yukawa couplings of the superpotential. This has important
effects both in the fermion mass matrix and in the scalar mass matrix. To
include the latter we need to know the mechanism of supersymmetry breaking.
We assume that it is broken in the hidden sector through a field $X$ with
non-vanishing vacuum expectation value for its F-term. Including this field
the leading terms in the K\"{a}hler potential for the matter fields are
given by 
\begin{eqnarray}
&K=\psi _{i}^{\dagger }\psi _{j}\left( \delta ^{ij}(c_{0}+d_{0}XX^{\dagger
})+\frac{\displaystyle{1}}{\displaystyle{M_f^{2}}}[\theta _{3}^{i\dagger
}\theta _{3}^{j}(c_{1}+d_{1}XX^{\dagger })+\theta _{23}^{i\dagger }\theta
_{23}^{j}(c_{2}+d_{2}XX^{\dagger })]+\right. &  \notag  \label{kahler} \\
&\left. +\Frac{1}{M_f^4}(\epsilon ^{ikl}\overline{\theta }_{3,k}
\overline{\theta }_{23,l})^{\dagger }(\epsilon ^{jmn}
\overline{\theta }_{3,m}\overline{\theta }_{23,n})
(c_{3}+d_{3}XX^{\dagger })\right) &
\end{eqnarray}%
where $\psi$ represents the $SU(2)$ doublet or the up and down singlets and 
accordingly $M_f=M_L, M_u, M_d$. Then $c_{i},d_{i}$ are ${\mathcal{O}}(1)$ 
coefficients in the different terms. After $SU(3)_{fl}$ breaking we obtain 
wave function and soft mass
corrections because the fields acquire non-zero vevs and F-terms.

\subsubsection{Wave function effects}

Due to the terms independent of $X,$ the matter fields do not have canonical
wave functions in the basis $\psi _{i}^{c},$ $\psi _{j}$ \cite{peddie}.
Rather, in the case of the $SU(2)$ doublets, they are given by%
\begin{eqnarray}
&&\left( 
\begin{array}{ccc}
\psi _{1}^{\dagger } & \psi _{2}^{\dagger } & \psi _{3}^{\dagger }%
\end{array}%
\right) \left( 
\begin{array}{ccc}
c_{0}+c_{3}a_{3}^{2}b_{23}^{2}/M_{L}^{4} & 0 & 0 \\ 
0 & c_{0}+c_{2}b_{23}^{2}/M_{L}^{2} & c_{2}b_{23}^{2}e^{-i\beta
_{3}}/M_{L}^{2} \\ 
0 & c_{2}b_{23}^{2}e^{i\beta _{3}}/M_{L}^{2} & 
c_{0}+c_{1}(a_{3}^{u\,2}+a_{3}^{d\,2})/M_{L}^{2}%
\end{array}%
\right) \left( 
\begin{array}{c}
\psi _{1} \\ 
\psi _{2} \\ 
\psi _{3}%
\end{array}%
\right)  \notag \\
&\equiv &\left( 
\begin{array}{ccc}
\psi _{1}^{\dagger } & \psi _{2}^{\dagger } & \psi _{3}^{\dagger }%
\end{array}%
\right) N^{\dagger }N\left( 
\begin{array}{c}
\psi _{1} \\ 
\psi _{2} \\ 
\psi _{3}%
\end{array}%
\right)
\end{eqnarray}%
where the messenger mass $M_{L}$ corresponds to heavy $SU(2)_{L}$ doublet
fields and the coefficients $c_{i}$ are of $O(1).$ There is an analogous
expression for the wave functions of the conjugate fields $\psi ^{c}$ with $%
M_{L}$ replaced by $M_{u,d}$ the mass of the ${u,d}$ $SU(2)$ singlet
messenger masses and $a_{3}^{u\,2}+a_{3}^{d\,2}$ replaced by $a_{3}^{u\,2}$ or 
$a_{3}^{d\, 2}$ for the up and down sectors respectively$.$ As discussed in 
\cite{king}, $M_{L}\succsim M_{u}> M_{d}$ is necessary if the expansion
parameter for the up quarks is to be smaller than that for the down quarks
and the Majorana mass matrix is to have a phenomenologically acceptable form.

The effect of these wave function corrections on the Yukawa couplings may be
determined by redefining the matter fields to obtain canonical kinetic
terms, $\psi ^{\prime }=N\psi .$ Consider first the contribution to the
observed mixing angles from the $\psi $ wave function effects. From 
\eq{expansion} we have $\frac{b_{23}^{2}}{a_{3}^{d\,2}}=\bar{\varepsilon}^{3}.$
In this case we have $\psi ^{\prime }=N\psi =R^{\dag }X^{-1}P\psi $ where 
\begin{eqnarray}
\label{cannonicalis}
X &=&\left( 
\begin{array}{ccc}
r_{1} & 0 & 0 \\ 
0 & r_{2} & 0 \\ 
0 & 0 & r_{3}%
\end{array}%
\right) =\left( 
\begin{array}{ccc}
1+\bar{\varepsilon}^{5}/2 & 0 & 0 \\ 
0 & 1+\bar{\varepsilon}^{4}/2 & 0 \\ 
0 & 0 & 1+\bar{\varepsilon}/2%
\end{array}%
\right)  \\
R &=&\left( 
\begin{array}{ccc}
1 & 0 & 0 \\ 
0 & 1 & \bar{\varepsilon}^{3} \\ 
0 & -\bar{\varepsilon}^{3} & 1%
\end{array}%
\right)  \\
P &=&\left( 
\begin{array}{ccc}
1 & 0 & 0 \\ 
0 & e^{i\beta _{3}/2} & 0 \\ 
0 & 0 & e^{-i\beta _{3}/2}%
\end{array}%
\right) 
\end{eqnarray}%
This is common to the up and the down sectors. As a result, if $X$ were the
unit matrix, the effect of $R$ would be to give a common contribution, $%
R^{\dag },$ to the matrices diagonalising the up and down quark mass
matrices, $V_{u}$ and $V_{d}.$ This can be absorbed in a redefinition of the
LH\ multiplets and clearly does not generate a CKM element. Thus the leading
contribution from these terms to the CKM matrix comes from the departure of $%
X$ from the unit matrix and is thus $O(\bar{\varepsilon})$ suppressed and
occurs at $O(\bar{\varepsilon}^{4}).$ Thus the mixing introduced in the $(2,3)
$ sector through the Kahler terms involving the fields $\psi $ is much
smaller than that coming from the superpotential \cite{glosi}. Turning to
the Kahler potential of the fields $\psi ^{c}$ sector, due to the ordering $%
M_{u}>M_{d}$ the mixing introduced is different for the up and the
down sectors and is larger in the down sector. Due to this there is a mixing
of $O(\bar{\varepsilon}^2)$ introduced in the $\psi ^{c}$ sector from the
Kahler potential \cite{glosi}. This mixing in the right handed sector can 
also modify the CKM mixing, however this effect is strongly suppressed. 
If the matrix $X$ relevant for the $\psi ^{c}$ sector is the unit matrix 
the effect vanishes. Including the
effect of the higher order terms responsible for making $X$ different from
the unit matrix we find a suppression by the factor 
$(m_{c}/m_{b})(-r_{2}^{2}+r_{3}^{2})$ times the mixing diagonalising the 
right handed sector $O(\bar{\varepsilon})$. Thus, this results in a 
contribution again of $O(\bar{\varepsilon}^{4})$ and much smaller than 
that coming from the superpotential \cite{glosi}.

One may readily check that the effects coming from the Kahler potential also
do not affect the MNS matrix to leading order.

\subsubsection{Soft mass corrections}

The K\"{a}hler potential Eq.~(\ref{kahler}) also gives corrections to the soft
SUSY breaking scalar masses generated by the vev for the F-term of the $X$
field. Redefining the matter fields to obtain the canonical kinetic terms 
we obtain the soft breaking mass matrices as, $M_{ij}^{2}=F_{a}F^{b}(%
\partial \log K_{ij})/(\partial \phi _{a}\partial \phi ^{b})$. In general
the coefficients $c_{i}$ will not be proportional to $d_{i}$ and therefore
diagonalising and normalising the kinetic terms will not diagonalise the
sfermion mass matrices. Then in this model, suppressing factors of order 1
due to the non-proportionality of $c_{i}$ and $d_{i}$, we have the
contribution to the scalar masses of the form 
\begin{equation}
M_{\tilde{f}}^{2}\simeq \left(\left( 
\begin{array}{ccc}
1 &  &  \\ 
& 1 &  \\ 
&  & 1%
\end{array}
\right) +\left( 
\begin{array}{ccc}
\varepsilon ^{2} & 0 & 0 \\ 
0 & \frac{b_{23}^{2}}{a_{3}^{f\,2}} & \frac{b_{23}^{2}}{a_{3}^{f\,2}}e^{i\beta
_{3}} \\ 
0 & \frac{b_{23}^{2}}{a_{3}^{f\,2}}e^{-i\beta _{3}} & 1%
\end{array}%
\right) \frac{a_{3}^{f\,2}}{M_f^{2}}\right)m_{0}^{2}
\label{soft1}
\end{equation}
where $m_0=|F_X|^2/M_{Planck}$ and $M_f=M_L,M_{u},M_{d}$ for the doublet 
squarks and the $u,d$ singlet squarks respectively.

\section{Phenomenological analysis of phases and $CP$ violation in the quark sector} 
\label{sec:quark}
The general structure of the Yukawa matrix and the soft breaking
mass matrices in the $SU(3)$ family symmetry models is given by 
Eqs.~(\ref{thematrix}) and (\ref{soft1}).
The parameters are fixed mainly by the quark masses and mixing angles
and in the fit to data of the quark textures the presence of phases plays an
important role. Furthermore $CP$ violation has been measured experimentally
in both the neutral kaon and neutral B sectors and we must reproduce
these measurements.
 
To address this we turn to a detailed discussion of the phenomenological fit. It is convenient to use the 
Wolfenstein parameterisation for the CKM matrix with parameters $\lambda$, $\rho$ and $\eta$. We have
\beq
\frac{|V_{ub}|}{|V_{cb}|}=\frac{\lambda}{c_\lambda}\sqrt{\rhobar^2+\etabar^2},\quad
\frac{|V_{td}|}{|V_{ts}|}=\frac{\lambda}{c_\lambda}\sqrt{(c_\lambda-\rhobar)^2 
+\etabar^2},\quad
|V_{us}|=\lambda
,
\eeq
for $\rhobar=c_\lambda\rho$ and $\etabar=c_\lambda\eta$, 
$c_\lambda=1-\lambda^2/2$.  The
experimental constraints are the semileptonic decays of B mesons,
whose charmless channel determines the ratio $|V_{ub}/V_{cb}|$, the CP
violation in the K system, from which we obtain the asymmetry
parameter $|\epsilon_K|$, the measurements of
$B^0_{d,s}-\bar{B^0_{d,s}}$ oscillations, from which the mass
differences $\Delta m_{B_d}$ and $\Delta m_{B_s}$ can be obtained, and
the CP asymmetries in various $B$ decays, from which $\sin(2\beta)$ is
obtained with $\beta$ one of the angles of the unitarity triangle.
We have updated the 2001 fit with the latest experimental information
{\footnote{The input data used in this fit are as given in in Appendix A, 
Table \ref{tab:experiment} and in Table 1 of 
\cite{romanino}.}}.
In this fit we have included the constraints $|V_{ub}/V_{cb}|$, 
$\Delta m_{B_s}$, $\Delta m_{B_d}$ and $|\epsilon_K|$. 
In the limit where we neglect all supersymmetric contributions to these
observables, the fitted values for $\rhobar$ and $\etabar$ are 
\bea
\label{rhoeta2003}
\rhobar=0.199^{+0.053}_{-0.049},\quad \etabar=0.328^{+0.037}_{-0.036},
\eea
We prefer to leave aside the constraints coming from the decay modes of B, 
given the present uncertainty regarding the decay mode $B\rightarrow 
\Phi K_S$.

Let us comment on how the phases and the mixing angle $s^d_{13}$ are fixed. For matrices with the hierarchy of the form of \eq{thematrix},
the magnitudes of the ratios $V_{td}/V_{ts}$ and $V_{ub}/V_{cb}$
depend only on two phases, $\Phi_2$ and $\Phi_3$ \cite{romanino},
\bea
\label{v31v32v13v23}
\frac{|V_{31}|}{|V_{32}|}=\frac{|V_{td}|}{|V_{ts}|}\approx
\left| s^d_{12}-e^{i\Phi_2}\frac{s^d_{13}}{s^Q_{23}} \right|\nonumber\\
\frac{|V_{13}|}{|V_{23}|}=\frac{|V_{ub}|}{|V_{cb}|}\approx
\left| s^u_{12}-e^{-i\Phi_3}\frac{s^d_{13}}{s^Q_{23}} \right|
\eea
with 
\bea
s^{Q}_{23}e^{-i\xi_s}=s^d_{23}
e^{-i(\gamma^d_{23}-\gamma^u_{23})}-s^u_{23},
\eea
$s^d_{ij}$ and $s^u_{ij}$ the rotations diagonalising the down and up
quark Yukawa textures respectively and $\gamma_{ij}^f= \mbox{Arg}(H_{ij}^f) = 
\mbox{Arg}(\sum_k m^f_{i k} m^{f\,*}_{j k})$.
The ratio $\frac{|V_{ub}|}{|V_{cb}|}$ is the  most sensitive 
to the presence of a phase in \eq{v31v32v13v23} due to the contribution 
coming from $s^d_{13}/s^Q_{23}$.
At first approximation, the phase $\Phi_1$ can be fixed by using the GST 
relation
\beq
\label{vus}
V_{us}\approx |s^d_{12}-e^{-i\Phi_1}s^u_{12}|=\left|\sqrt{\frac{m_d}{m_s}}-
e^{-i\Phi_1}\sqrt{\frac{m_u}{m_c}} \right|,
\eeq
where we have used that $s^{u}_{12}=\sqrt{\frac{m_u}{m_c}}$ and 
$s^d_{12}=\sqrt{\frac{m_d}{m_s}}$, which applies for the structure of the mass 
matrix in \eq{thematrix}.
The ratios of masses $m_u/m_c$ and $m_d/m_s$ 
can be written in terms of ratios  $m_u/m_d$ and the parameter 
$Q$, which are better determined from chiral perturbation theory 
\cite{leutwyler} (see table 3 in \cite{romanino}) and the ratio $m_s/m_c$ can 
be obtained from the $\overline{MS}$ values at a common scale 
\cite{toni,charm,arcadi}. This information gives us
\beq
\label{ratmdms}
\sqrt{\frac{m_d}{m_s}}=0.228 \pm 0.005, ~~~~~~ 
\sqrt{\frac{m_u}{m_c}}=0.054 \pm 0.011.
\eeq
Comparing \eq{vus} with the experimental value of $V_{us}=0.2240 \pm 0.0036$, 
we obtain $\Phi_{1}=\pm\left( 1.3 \pm 0.6 \right)$ which corresponds in 
degrees to $\Phi_1=  \pm \left( 74^0\pm 34^0 \right)$.

From here, we see that non zero phases are needed to fit quark masses 
and CKM mixings, even before considering $CP$ violation observables. Taking 
into account that $\Phi_1$ is the main phase contributing to the Jarlskog 
invariant, as we will see below, this implies that its value is in the 
upper range of this interval. 

However in order to fix appropriately the phases and the mixing angle 
$s^d_{13}$, we need to give a common solution to the system of the two 
equations appearing in \eq{v31v32v13v23} and to \eq{vus} for the variables 
$\Phi_2$, $\Phi_1$ and $s^d_{13}$, and we need to use a constraint on 
$\Phi_3$ in order to write the equations just in terms of the above 
variables. For matrices of the form \eq{fit}, $\Phi_1$ $\Phi_2$ and 
$\Phi_3$ are given by,
\bea
\label{Phis}
\Phi_1&\approx& (\gamma^{u}_{12}-\gamma^{d}_{12})-(\gamma^{u}_{22}-\gamma^{d}_{22})\nonumber\\
\Phi_2&\approx& (\gamma^{d}_{13}-\gamma^{d}_{23}-\gamma^{d}_{12}+\gamma^{d}_{22})-\frac{s^u_{23}}{s^d_{23}}\sin(\gamma^d_{23}-\gamma^u_{23})-\sin(\phi^d) s^d_{23}c^d_2\nonumber\\
\Phi_3&\approx & (\gamma^{d}_{13}-\gamma^{d}_{23}-\gamma^{u}_{12}+\gamma^{u}_{22})+\frac{s^u_{23}}{s^d_{23}}\sin(\gamma^d_{23}-\gamma^u_{23})-\sin(\phi^d)s^d_{23}c^d_2,
\eea
with $\phi^f=\gamma_{13}^f-\gamma_{12}^f-\gamma_{23}^f$, the coefficients 
$c^f_{1,2}$ are of order 1, and as defined above  $\gamma_{ij}^f= 
\mbox{Arg}(\sum_k m^f_{i k} m^{f\,*}_{j k})$. Thus we can write 
\beq
\label{deltas}
\Phi_3\approx (\Phi_2-\Phi_1).
\eeq
Hence the ratios Eqs.~(\ref{v31v32v13v23},\ref{vus}) of the CKM matrix, 
can be expressed just in terms of two phases, $\Phi_1$ and $\Phi_2$
\bea
\frac{|V_{td}|}{|V_{ts}|}&\approx&
\sqrt{\frac{m_d}{m_s}+C^2- 2C\sqrt{\frac{m_d}{m_s}}\cos(\Phi_2-\Phi_1)}\nn\\
\frac{|V_{ub}|}{|V_{cb}|}&\approx&
\sqrt{\frac{m_u}{m_c}+C^2-2C\sqrt{\frac{m_u}{m_c}}\cos(\Phi_2)}
\label{VsC}
\eea
for $C=c\bar \varepsilon^0$ (with $c$ the ${\cal{O}}(1)$ coefficient
in \eq{fit}), for $\bar \varepsilon^0=0.15\pm 0.01$
{\footnote{The
quantities with superscript 0 refer to the quantities appearing in the
parameterisation of Eq.~(25) in \cite{romanino}.}}.Solving for the three equations, Eqs.~(\ref{vus},\ref{VsC}) we have the following solution, requiring $c$ to be real and to satisfy $0 \leq c \leq 1$,
\bea
\Phi_1&=1.57\pm 0.02 &  \longrightarrow \quad
c=0.28\pm 0.09,\quad \Phi_2=0.8\pm 0.3 
\label{fit03}
\eea
Notice that the inclusion of the constraints in \eq{VsC} modifies
substantially the mean and uncertainty of $\Phi_1$ from the result we
obtained just from \eq{vus}. We have included {\it all} the
uncertainties of the parameters 
involved in Eqs.~(\ref{v31v32v13v23}) and (\ref{vus})\footnote{We can
compare these solutions with those of the fit of \cite{romanino} 
where $\rho, \eta$ and $\lambda$ are different: $c=0.5,\quad
\Phi_2=-24^0$, $c=0.95,\quad\Phi_2=-(58\pm 5^0)$ for $\Phi_1=+90^0$
and $\Phi_1=-90^0$ respectively.}.

\subsection{Comparison with the $SU(3)$ texture.}
These values must be compared with the predicted values in our model 
from \eq{Phis} and the leading order texture in \eq{thematrix},
\bea
\Phi_1 
= 2 (\alpha_u - \alpha_d)
\eea
\beq
\Phi_2\approx - \frac{s^u_{23}}{s^d_{23}}\sin(\gamma^d_{23}-\gamma^u_{23})
\label{expPhi2}
\eeq
where we used the fact that, at leading order, 
$\gamma^d_{23} = \beta_3 - 2 \chi$, $\gamma^d_{12} = \delta^d$, 
$\gamma^d_{13} = \delta^d + \beta_3 - 2 \chi$, $\gamma^u_{23} = \beta_3$,
$\gamma^u_{12} = \delta^u$ and $\gamma^u_{13} = \delta^u + \beta_3$. So, 
we obtain $\Phi_2 \approx 0.11$ assuming that the phase 
$\gamma^d_{23}-\gamma^u_{23} = \delta^d - \delta^u - 2 \chi$ is ${\cal O}(1)$. 
Therefore, from the leading order texture in \eq{thematrix} 
we can only be compatible at the 3$\sigma$ level with the required value from 
the SM fit which corresponds to $\Phi_2=0.771^{+0.308}_{-0.316}$.

However, there are several other contributions which can affect this result.
In particular there are subleading contributions in the 
$\varepsilon$ expansion to the different elements of the Yukawa texture 
and also there may be significant supersymmetric contributions to SM 
processes which can affect the fit to $\Phi_2$.
\subsection{Subleading contributions}
 In the case of Model II analysed later 
\cite{king2}, we have an additional contribution to $Y^d_{23}$ such that,
\bea
Y^d_{23}~ =~ Y^d_{32}~= ~\bar \varepsilon^2~ e^{i \beta}~ \Frac{\Sigma_d}
{|a_3^d|^2} + \tilde c~ \bar \varepsilon^{2.5} e^{i \theta_S}
\eea
with $\tilde c$ an unknown coefficient order 1. Then we have,
\bea
\Phi_2\approx ~\tilde c ~\Frac{|a_3^d|^2}{\Sigma_d}~ \bar \varepsilon^{0.5} 
\left( 1 - \frac{X_2}{X_1 + X_2} \right) ~\sin (\beta_3 - \theta_S) + 
{\cal{O}}(\frac{s^u_{23}}{s^d_{23}})
\label{Phi2II}
\eea
Therefore, in this case we have a contribution of order 
$\sqrt{\bar \varepsilon} \simeq 0.4$ which can accommodate the result of 
the fit within 1$\sigma$ of $\Phi_2$.

\subsection{SUSY contributions}
Another possibility is that there are significant SUSY contributions to 
SM processes that have not
been taken into account in the SM fit and can play an important role 
in determining
the value of $\Phi_2$. A full computation of the complete 
SUSY contributions is beyond the scope of this paper, but we can  
estimate the size of the new SUSY contributions and fit a modified
standard model triangle.

Using Eqs.~(\ref{thematrix}) and (\ref{soft1}) we can obtain the 
sfermion mass matrices in the so called SCKM basis with the correct
phase convention in the CKM mixing matrix. The SCKM basis is the
basis where we rotate the full supermultiplet, both quarks and
squarks, simultaneously to the basis where quark Yukawas are diagonal
and real. Then, we must also ensure that the   phases in
the CKM matrix are located in the $(1,3)$ and $(3,1)$ elements
up to order $\bar \varepsilon^4$. This is done explicitly in Section
\ref{sec:SUSY} and here we simply use these results. 
To analyze flavour violating constraints at the electroweak scale, the
model independent Mass-Insertion (MI) approximation is advantageous
\cite{MI}. A Mass Insertion is defined as $\delta_{ij}^f
\equiv \Delta^f_{ij}/m_{\tilde{f}}^2$; where $\Delta^f_{ij}$ is the
flavour-violating off-diagonal entry appearing in the $f = (u,d,l)$
sfermion mass matrices in the SCKM basis and $m_{\tilde{f}}^2$ is 
the average sfermion mass. In addition, the MI are further sub-divided into
LL/LR/RL/RR types, labeled by the chirality of the corresponding SM
fermions. The different Mass Insertions \cite{MI} at the electroweak scale
relevant for $\varepsilon_K$ are,   
\bea
\label{predeps}
\sqrt{|\mbox{Im} (\delta_{RR}^d)^2_{12}|}  = 6.8 
\times 10^{-4} \sin \Phi_1\\
\sqrt{|\mbox{Im} (\delta_{RR}^d)_{12} (\delta_{LL}^d)_{12}|} = 
2 \times 10^{-4} \sin \Phi_1\nn
\eea 
These values are only order of magnitude estimates
representing the maximum possible value for these entries and they can
be reduced by factors order one in the redefinition of the fields to
get the canonical kinetic terms as explained in Section \ref{sec:SUSY}.
This must be compared with the phenomenological MI bounds \cite{MI},
\bea
\sqrt{|\mbox{Im} (\delta_{RR}^d)^2_{12}|} \leq 3.2 \times 10^{-3} \nn\\   
\sqrt{|\mbox{Im} (\delta_{RR}^d)_{12} (\delta_{LL}^d)_{12}|}
\leq 2.2 \times 10^{-4}
\eea
These bounds correspond to average squark masses of 500~GeV and they scale
as $m_{\tilde q}~$(GeV)/500 for different squark masses.  
Then, for $m_{\tilde q}\simeq 500$ GeV, it is clear that we can assume a 
sizeable SUSY contribution to $\varepsilon_K$ that can reach a $20\%$ of 
the experimental value. For larger squark masses a $20\%$ contributions 
is still possible if we consider simultaneously $(\delta_{RR}^d)_{12}$ 
and $(\delta_{LL}^d)_{12}$. 

To take into account of a possible SUSY contribution up to a $20\%$ of the 
experimental value, we enlarge the error on $\varepsilon_K$ to $20\%$ of the 
mean value. The resulting fit is given in Figure \ref{fig:fcls2003s20pA} giving
\bea
\bar{\rho}=0.212^{+0.046}_{-0.048},\qquad \bar{\eta}=0.348^{+0.04}_{-0.042},
\eea
which in turns imply different values for the parameters $c$, $\Phi_2$:
\bea
\Phi_1=1.58\pm 0.02,\quad
c=0.297^{+ 0.082}_{-0.079},\qquad \Phi_2=0.830^{+0.310}_{-0.320}
\eea
Although here the errors on $\Phi_2$ have increased, so the mean, which again would be compatible with the result of \eq{expPhi2} only at 3$\sigma$ level.
This shows that the favoured solution prefers a SUSY contribution of 
opposite sign to the SM contribution.

However it is instructive to determine the implication of a
supersymmetric contributions to $\epsilon_K$ which {\it has} positive 
sign with respect to the SM contribution, such that the SM
value of $\epsilon_K$ is only $80\%$ the mean of the experimental
value of $\epsilon_K$. Fixing this a fit to the revised data gives
\bea
\rhobar=0.228^{+0.057}_{-0.059},\quad \etabar=0.306^{+0.050}_{-0.052}\nn\\
\Phi_1=1.57\pm 0.02 ,\quad c=0.336^{+0.133}_{-0.133},\quad \Phi_2=0.599^{+0.350}_{-0.334}
\label{eq:cphi2epskm20M}
\eea
Although in this case the quality of the fit is somewhat smaller than
in the previous fit, now the model I is compatible within 2$\sigma$ of
the result of the fit for $\Phi_2$ -\eq{eq:cphi2epskm20M}, while the 
model II is compatible at 1$\sigma$ of the fitted value.

\subsection{The Jarlskog invariant}
It is instructive also to calculate the Jarlskog invariant in the CKM
matrix which can be computed from 
\beq
\label{cpviinv}
J_{\rm CP}={\rm Im}\{V^*_{32}V^*_{21}V_{22}V_{31}\}
\eeq
where $V$ is the CKM matrix, which depends on the sines of the mixing
angles, $ \sin^f(\theta_{ij})\equiv s^f_{ij}$.  In this case we have
\begin{equation}
\label{smallmix}
s_{12}^{f}\approx \frac{|m^f_{12}+\frac{m^f_{21}m^f_{11}}{m^f_{22}}|}
{|m^f_{22}|},\qquad s_{23}^f\approx
\frac{|m^f_{23}+\frac{m^f_{32}m^f_{22}}{m^f_
{33}}|}{|m^f_{33}|},\qquad s_{13}^{f}\approx \frac{
|m^f_{13}+\frac{m^f_{31}m^f_{11}}{m^f_{33}}|}{|m^f_{33}|},
\end{equation}
which to leading order are given by 
\beq 
\label{order}
s^f_{12}=O(\epsilon),\quad
s^f_{23}=O(\epsilon^2), \quad 
s^f_{13}=O(\epsilon^3), 
\eeq 
with $\epsilon=\bar \varepsilon$ for $f=d$-quarks and
$\epsilon=\varepsilon$ for $f=u$-quarks and $\bar \varepsilon\gg
\varepsilon$. Thus $J_{\rm CP}$ acquires the form 
\bea
\label{invJcp}
J_{\rm CP}&\approx&{\rm Im}\{s^{Q}_{23}\left[s^d_{12}s^u_{12}s^{Q}_{23}
e^{i\Phi_1}-s^d_{12}s^d_{13}e^{i\Phi_2}-s^u_{12}s^d_{13}e^{i\Phi_3}
\right]\}\nonumber\\
s^{Q}_{23}e^{-i\xi_s}&=&s^d_{23}e^{-i(\gamma^d_{23}-\gamma^u_{23})}-s^u_{23}
\eea 
with $s^{Q}_{23}$ real.

In a particular parameterisation of the CKM mixing matrix we have a single 
unremovable $CP$ violating phase which is responsible for $CP$ violation 
through this invariant. In the  PDG ``standard'' parameterisation \cite{PDG}
the Jarlskog invariant is $J_{\rm CP} = s_{12} c_{12} s_{23} c_{23}s_{13} 
c_{13}^2 \sin \delta$, with $c_{ij}, s_{ij}$ the cosine or sine of a 
rotation in the $(i,j)$ plane. Using the measured values of the mixing angles 
at the electroweak scale we obtain  $J_{\rm CP} = 3.3 \times 10^{-5}~ \sin 
\delta \simeq (2.8 \pm 0.4) \times 10^{-5}$, with $\delta = 1.02 \pm 0.22$.
In our model the Jarlskog invariant in \eq{invJcp} must reproduce this value up
to a correction in the phase due to the relatively small SUSY contributions 
to the unitarity triangle fit (see Figure \ref{fig:fcls2003s20pA}).
We must remember that $\Phi_1$ is nearly maximal, while $\Phi_2$ is at most 
order $\sqrt{\bar \varepsilon}$ or smaller. Using this together with the 
fitted values of the coefficients from \eq{fit} this implies that the 
Jarlskog invariant is dominated by the first contribution in 
\eq{invJcp}. If we use the values of the angles at $M_{GUT}$ we obtain,
\bea
\label{JcpGUT}
\left.J_{\rm CP}\right|_{M_{GUT}}&\approx& a^2~ b~b^\prime \bar \varepsilon^5 
\varepsilon \sin \Phi_1 \simeq 1.44 \times 10^{-5} \sin \Phi_1
 \eea
where we used that $b^\prime = b$ as the (1,2) entries are not affected by 
the vev of $\Sigma$ breaking $SU(2)_R$. However, this is the value of the 
invariant at $M_{GUT}$ and we need to evolve it to $M_W$ to compare with the 
experimental results. We can use the RGE equations for the Jarlskog invariant 
in \cite{Olechowski:1990bh,barger} and we obtain,
\bea
\label{JcpMW}
\left.J_{\rm CP}\right|_{M_{W}}&\simeq& \chi^{-2} 
\left.J_{\rm CP}\right|_{M_{GUT}} \simeq 2.9 \times 10^{-5} \sin \Phi_1
\eea
where $\chi = (M_{GUT}/M_Z)^{-Y_t^2/(16\pi^2)} \simeq 0.7$ 
\cite{barger,romanino}. By comparison with the experimental result for the Jarlskog invariant we again conclude that  
$\Phi_1$ must be near maximal.

\section{Neutrino masses, mixings and CP violation}
\label{sec:neutrino}
As we have seen in the previous Sections the quark Yukawa couplings 
fix many of the features of our $SU(3)_{fl}$ model, but we still have
some freedom to build the right- handed Majorana neutrino mass matrices.
To do this we must take into account that this matrix violates lepton number
by two units. As discussed in Appendix C we assume this breaking comes from a sneutrino 
triplet, $\lambda$, and antitriplet, $\bar{\lambda}$ with vevs $\lambda = (0,0,c)$ and 
$\bar{\lambda} = (0,0,c)$. Then the dominant Majorana mass comes from 
\bea
W_M = \Frac{1}{M} \psi^c_i \lambda^i \psi^c_j \lambda^j \, + \,\dots
\eea
 Any other term in the superpotential will have to include this 
set of fields because $\lambda$ is the only term violating lepton number 
by 1 unit. Therefore all other terms are obtained by combining this
term with a neutral combination of fields and reordering the $SU(3)$
indices. The allowed terms will depend on the global symmetries
that must reproduce the correct quark and charged lepton Yukawa textures and at the same time 
give rise to a suitable Majorana mass matrix. As mentioned in Section I
the choice here is not unique and we have several possibilities.
Here we will give two concrete examples which reproduce the correct 
Yukawa textures and give an acceptable neutrino mass matrix. 
Since $CP$ is spontaneously broken only in the flavour sector all the phases 
in the lepton sector are a function of the flavon phases. Consequently
there are predictions for the $CP$ violating phases in neutrino physics.
We will illustrate this in the two examples given below.

\subsection{Model I}

In this example we need two additional global symmetries to obtain 
the correct Yukawa textures. These include a continuous 
R-symmetry which can play the role of the Peccei-Quinn symmetry \cite{peccei}
to solve the strong CP problem and a discrete symmetry $Z_{3}\otimes Z_{5}$, 
under which the fields transform as shown in Table \ref{tab:higgs}.
\begin{table}[tbp] 
\begin{tabular*}{1.00\textwidth}{@{\extracolsep{\fill}}|c|c c c c c c c c 
c c|}
\hline
~${\bf Field}$ & $\psi$&$\psi^c$&$\lambda$&$H$&$\Sigma$&
$\theta _{3}$ & $\theta _{23}$ & $\overline{\theta _{3}}$
&$\overline{\theta _{23}}$&$\overline{\theta _{2}}$~~~ \\ \hline 
~${\bf SU(3)}$ &${3}$ &${3}$&${3}$ &$1$&$1$& $\overline{3}$ &
$\overline{3}$ &${3}$ &${3}$ &${3}$~~~ \\
~${\bf R }$& $1$ &  $1$ & $-2$ & $2$ &  $0$ & $-3$ & $-3$ &  $3$ 
&  $-6$ & $12$ ~~~ \\
~${\bf Z_{5}}$& $0$ &  $0$ &  $0$ & $0$ &  $1$ & $0$& $2$ & $2$& 
$2$ & $1$~~~\\
~${\bf Z_{3}}$& $0$ &  $0$ &  $0$ & $0$ &  $1$ & $0$& $1$ & $2$& 
$0$ & $2$~~~\\
\hline
\end{tabular*}
\caption{Transformation of the matter superfields under the family
symmetries.}
\label{tab:higgs}
\end{table}
\subsubsection{Dirac structure}
Using these charges we can build the effective Yukawa couplings as,
\bea
W_Y = H \psi_i \psi^c_j \left[ \theta_3^i \theta_3^j \, + \, 
\theta_{23}^i \theta_{23}^j \Sigma \, + \, 
\left( \epsilon^{i k l} 
\overline{\theta}_{23,k} \overline{\theta}_{3,l} 
\theta^j_{23} \, + \,\epsilon^{j k l} 
\overline{\theta}_{23,k} \overline{\theta}_{3,l} 
\theta^i_{23} \right)\left(\theta_{23} \overline{\theta _{3}}\right)
\right. \nn \\ 
\left. \epsilon^{i j k} \overline{\theta}_{23,k} 
\left(\theta_{23} \overline{\theta _{3}}\right)^2 \, + \,
 \epsilon^{i j k} \overline{\theta}_{3,k}  
\left(\theta_{23} \overline{\theta _{3}}\right)  
\left(\theta_{23} \overline{\theta _{23}}\right)\,+ \,  
\theta_3^i \theta_{23}^j \left(\theta_{3} 
\overline{\theta _{3}}\right)^4  \right]
\label{YukawaWI}
\eea
and this reproduces the structure in \eq{YukawaW0}, apart from
the last term that will give a subdominant contribution to the $(2,3)$ and
$(3,2)$ elements of the Yukawa matrix.
The neutrino Yukawa matrix is given by \eq{YukawaWI} 
with $\Sigma_\nu = 0$,
\bea
\label{thematrixI}
Y^{\nu}=\left(
\begin{array}{ccc}
0& \varepsilon^3e^{i\delta^u}\, X_1   
& \varepsilon^3e^{i(\delta^u + \beta_3)}\, X_2 \\
...& 0 &
\eta \varepsilon^{2.75} e^{i 4 \alpha_u}\\
...&... &1
\end{array}
\right) \frac{|a^u_3|^2}{M_u^2} 
\label{xxx}
\eea
with $\eta$ an order 1 coefficient.
\subsubsection{Majorana structure}
In a similar way we can calculate the structure of the Majorana matrix.
As explained above we simply have to look for the neutral combinations 
under the extra symmetries. We have two neutral combinations
of fields that will be relevant: $\left(\theta_{23} \overline{\theta _{3}}
\right) \left(\theta_{3} \overline{\theta_{3}}\right)^3$
and $\left(\theta_{23} \overline{\theta _{3}}\right)^3 
\left(\theta_{3} \overline{\theta_{3}}\right) \overline{\theta_{3}}_i
\overline{\theta_{3}}_j \overline{\theta_{23}}_k$ (the combination
$\left(\theta_{3} \overline{\theta_{3}}\right)^7 \overline{\theta_{3}}_i
\overline{\theta_{3}}_j \overline{\theta_{23}}_k$ only contributes
to the element (1,3) and has a small impact on neutrino mixings). 
So, we obtain the different entries in the Majorana matrix from,
\bea
\label{MajI}
W_M = \Frac{\psi^c_i  \psi^c_j}{M} \left[ \lambda^i \lambda^j \, + \,
 \theta_3^i  \theta_{23}^j \left(\lambda \overline{\theta _{3}}\right)^2 
\left(\theta_{3} \overline{\theta_{3}}\right)^2 \, + \, 
\theta_{23}^i \theta_{23}^j \left(\lambda \overline{\theta _{3}}\right)^2 
\left(\theta_{3} \overline{\theta_{3}}\right)^6  \right. \\
\left.
 \,+ \, \epsilon^{i k l} 
\overline{\theta}_{23,k} \overline{\theta}_{3,l} 
\theta^j_{3}\left(\lambda \overline{\theta _{3}}\right)^2
\left(\theta_{3} \overline{\theta _{3}}\right)^6 \,+ \,
 \epsilon^{i k l} 
\overline{\theta}_{23,k} \overline{\theta}_{3,l} 
\theta^j_{23}\left(\lambda \overline{\theta _{3}}\right)^2
\left(\theta_{3} \overline{\theta _{3}}\right) 
\left(\theta_{23} \overline{\theta _{3}}\right)^2
\right] \nn
\eea
and assuming $b_{23}/M_u = \varepsilon$ and $a_3/M_u = \varepsilon^{0.25}$, 
this gives a Majorana matrix of the form,
\beq
\label{Mmatrix}
M_R=\left(
\begin{array}{ccc}
0& \lambda \varepsilon^{5.75} e^{i \delta^\prime}
& \varepsilon^{5} e^{i(\delta^\prime + 3 \alpha_u - 2 \beta_3)} \\
...&\eta_2\varepsilon^{5.5} e^{i(8 \alpha_u)} & \eta_1 
\varepsilon^{2.75} 
e^{i(4 \alpha_u)}\\
...&... &1
\end{array}
\right) \frac{c^2}{M} 
\eeq
with $\delta^\prime = (6 \alpha_u + 2 \beta_3 +\beta^\prime_2)$ and 
$\lambda$, $\eta_1$, $\eta_2$, ${\cal{O}}(1)$ coefficients.

Using Eqs (\ref{xxx}) and (\ref{Mmatrix}), the form of the 
low energy neutrino mass matrix  is determined in Appendix B, c.f. 
Eq (\ref{chifin}), together with the corresponding neutrino
masses and mixing angles. 

For a reasonable choice of the parameters we can easily obtain the observed masses and mixing angles.
Choosing $X_1=1.4$, $X_2 = 2$, $\eta=1.5$, $\lambda=0.34$, 
$\eta_2=0.75$ and  $\eta_1=1/3$ and neglecting charged lepton mixings,
we obtain for the neutrino mixing matrix
\bea
R^{\nu}=\left(
\begin{array}{ccc}
-0.78&0.58&0.22\\
-0.52& -0.42 &-0.74\\
0.34 & 0.69&-0.64
\end{array}
\right)
\eea
with eigenvalues $m_1 \simeq 0.007 \Frac{v_2^2}{M_3}$, 
$m_2 \simeq 0.12 \Frac{v_2^2}{M_3}$ and  $m_1 \simeq 1.02\Frac{v_2^2}{M_3}$, 
for $M_3$ the heaviest right-handed neutrino Majorana eigenvalue.
The small contributions to the mixing angles coming from the charged 
lepton sector,  \eq{chargedrotation} and 
\eq{chipp}, are only significant for the CHOOZ angle 
\cite{Velasco-Sevilla:2003gd,King:2002nf},
\bea
\sin \theta_{13} \simeq \left|~ \sin \theta_{13}^{\nu_L} - 
\sin \theta_{23}^{\nu_L} ~s_{12}^{l}~ e^{i \varphi_L}~ \right|
\eea 
From here we see that the CHOOZ angle can be reduced or enhanced depending on the
phase $\varphi_L$.

Regarding $CP$ violation in the neutrino sector, we have three different
observables, the MNS phase measurable in neutrino oscillations, 
the leptogenesis phase and the neutrinoless double beta decay phase.

In first place let us determine the MNS phase. First we define an
hermitian matrix $H^\nu_{ij} = \sum_k\chi^\prime_{ik} \chi_{jk}^{\prime\,*}$
with $\chi^\prime$ the low energy effective neutrino mass matrix
defined in Appendix \ref{sec:neutmamix}, 
\eq{chifin}.
We calculate the MNS phase through the invariant Im
$(H^\nu_{12}H^\nu_{23} H^\nu_{31})$ in the basis in which the charged
leptons are diagonal \footnote{Note that after some simple algebra, 
using the unitarity of the MNS mixing matrix, Im~$(H^\nu_{12}H^\nu_{23} 
H^\nu_{31}) = (m_{\nu_3}^2 - m_{\nu_1}^2)(m_{\nu_3}^2 - m_{\nu_2}^2)
(m_{\nu_2}^2 - m_{\nu_1}^2)~ \mbox{Im}~(V_{22} V_{23}^* V_{33} V_{32}^*)$.}. 
Then we notice that we can factor out the
phases $\tilde P_i \tilde P_j^{-1}$ from $H^\nu_{ij}$ and these phases 
cancel exactly in the invariant. Then the only remaining phase
up to order $\varepsilon^{2.75}$ is $\varphi_L = -2 \alpha_d - 2
\alpha_u+  \beta_3$. In this way, we can calculate the invariant from
$\chi^{\prime\prime}=\tilde{P}^{-1} \chi^{\prime} \tilde{P}^{-1}$,
which from \eq{chifin} can be written at leading order in $s_{12}^l$, as 
\beq 
\label{chipp}
\chi^{\prime \prime} \simeq \left(
\begin{array}{ccc} 
\chi_{11}-2\chi_{21}s^l_{12}e^{i\varphi_L}& \chi_{12}-(\chi_{22}e^{i\varphi_L}-
\chi_{11}e^{-i\varphi_L})s^l_{12}&\chi_{13}-\chi_{23}s^l_{12}e^{i\varphi_L}\\
...&\chi_{22}+2\chi_{12}s^l_{12}e^{-i\varphi_L}& 
\chi_{23}+\chi_{13}s^l_{12}e^{-i\varphi_L}\\
...&...&\chi_{33}
\end{array}
\right), \eeq 
for $s^l_{12}\approx \bar \varepsilon X_1/\Sigma_2$.  Then
the $H^{\prime\nu}_{ij}$ elements (as computed from
$\chi^{\prime\prime}$) can be written as follows 
\bea
H^{\prime\nu}_{12}&=&A_{12} + s^l_{12} e^{ i\varphi_L} ( |\chi_{11}|^2 -
|\chi_{22}|^2 + |\chi_{13}|^2 - |\chi_{33}|^2)\nn\\
H^{\prime\nu}_{31}&=&A_{31}- s^l_{12}e^{-i\varphi_L} A_{32} \nn\\
H^{\prime\nu}_{23}&=&A_{23}+ s^l_{12}e^{-i\varphi_L} A_{13} \eea where
we have defined $A_{ij} = \sum_{a=1,3} \chi_{i a} \chi_{j a}^*$ and we must 
remember that in this case $A_{ij}$ are real up to order 
$\varepsilon^{2.75}$. Then
\bea H^\nu_{12}H^\nu_{23} H^\nu_{31}\approx A_{12}A_{31}A_{23}+
s^l_{12}\left[ A_{12} (A_{31}^2 - A_{23}^2) e^{- i \varphi_L} + \right.
\nn \\
\left. A_{31} A_{23} (|\chi_{11}|^2 - |\chi_{22}|^2 + |\chi_{13}|^2 - 
|\chi_{33}|^2) 
e^{i \varphi_L}
\right] + \dots 
\label{invariant}
\eea 
where there can be higher powers in $s^l_{12}e^{\pm i\varphi_L}$, although
it is enough to keep the powers shown.
Due to the hierarchies in the elements $\chi_{ij}$, \eq{chiinterm}, we
have that $\chi_{1 i} <<
\chi_{22},\chi_{23},\chi_{32},\chi_{33}$. Then
we can approximate, 
\bea
\rm{Im}(H^\nu_{12}H^\nu_{23} H^\nu_{31})\approx 
\rm{Im}(A_{12}A_{23} A_{31}) +
 \left[\chi_{12} \chi_{23}^2 \left(\chi_{22} \chi_{33} (
\chi_{22} - \chi_{33}) + \chi_{23}^2 (\chi_{22} + \chi_{33})\right) +
\right. \nn\\ 
\left.
\chi_{13} \chi_{23} \chi_{22} \left( \chi_{23}^2 (\chi_{33} - \chi_{22}) +
\chi_{22} \chi_{33} ( \chi_{22} + \chi_{33} )\right)\right]~s^l_{12} 
\sin\varphi_L
\label{MSN}
\eea
This expression is completely general and includes the effect of non-negligible 
charged lepton mixings. In fact, we have 
$A_{12} A_{23} A_{31} = {\cal{O}}(\chi_{1a}^2 \chi_{bc}^4)$ with 
$a,b,c= 2,3$ while the square braket in \eq{MSN} is 
${\cal{O}}(\chi_{1a} \chi_{bc}^5)$. Therefore, it follows that the 
charged lepton phase will always contribute at leading order if 
$s^l_{12} \gsim \chi_{1a}/\chi_{ab}$ which in fact is usually satisfied 
in a model where charged lepton and down quark yukawas are related.
With the exception 
of the CHOOZ angle, the neutrino mixing 
angles are not affected by charged lepton mixing angles.   It is important to emphasize that there is no equivalent result when considering leptonic phases
\cite{King:2002qh,sacha}. As we see above, a relatively small charged 
lepton mixing is enough to affect at leading order the MNS phase. 

In our particular case, taking into account that $A_{12} A_{23} A_{31}$ is real
and replacing the elements
of the effective neutrino mass matrix, \eq{chiinterm}, in the case
$\eta = \eta^\prime$, we obtain,
\bea
\rm{Im}(H^\nu_{12}H^\nu_{23} H^\nu_{31})\approx s^l_{12}~\sin\varphi_L
\left(1 - \frac{\rho}{\lambda^2}~ c_{12}^2 X_2^2 \right) 4~
\frac{\rho^5}{\lambda^{10}}~ s_{12}~ c_{12}^9 (1 + s_{12}^2)~ X_1^6 X_2
~(X_1^2 + X_2^2)  
\label{MSNfin}
\eea
Provided, as is generally true, this imaginary part does not vanish the observable MNS phase is
given by this term, $\varphi_L = -2 \alpha_d - 2 \alpha_u+  \beta_3$.

The leptogenesis $CP$ asymmetry is of the form 
\bea
\label{leptogen}
\epsilon_1 = - \Frac{3}{16 \pi (Y^{\nu \dagger} Y^\nu)_{11}} \sum_{i\neq 1}
\mbox{Im}\left[(Y^{\nu \dagger} Y^\nu)^2_{1i}\right] \Frac{M_1}{M_i}
\eea
where the neutrino Yukawa matrices are given in the basis of diagonal 
Majorana masses, \eq{dirac'}. In this case, it is easy to see that 
the diagonal phase matrix cancels; in particular the phases coming from 
charged leptons do not contribute. Then the phase relevant for 
leptogenesis is given by
$\varphi =  4 \alpha_u + \beta_3$.

Finally, the neutrinoless double beta decay phase is
simply the relative phase between the two dominant contributions of 
$\chi^\prime_{11}$ in the basis of diagonal charged lepton masses. 
This time from \eq{chifin} we obtain,
\bea
(\chi^\prime)_{11} = \tilde P_1^2 \left(\chi_{11} - 2 \bar \varepsilon 
\frac{X_1}{\Sigma} \chi_{12} e^{-i \varphi_L} \right)
\eea
and if we remember that $\chi_{ij}$ are real up to order 
$\varepsilon^{2.75}$ we obtain the result that the neutrinoless double beta decay
phase is also $\varphi_L$ i.e. it coincides with the MNS phase.

\subsection{Model II}

\bigskip {\small 
\begin{table}[tbp]
\begin{tabular*}{1.0\textwidth}{@{\extracolsep{\fill}}|c|cccccccccccc|}
\hline
${\bf Field}$ &$ \psi$&$\psi^c$& $\lambda$&$\overline{\lambda}$&
$H$&$\Sigma$& $S$& $\theta _{3}$ & $\theta _{23}$ & $\overline{\theta _{3}}$
&$\overline{\theta _{23}}$&$\overline{\theta _{2}}~~$ \\ \hline
${\bf SU(3)}$ & ${\bf 3}$ & ${\bf 3}$ & ${\bf \overline{3}}$ & ${\bf
3}$ & ${\bf 1}$ &${\bf 1}$&${\bf 1}$ &${\bf \overline{3}}$ &
${\bf \overline{3}}$&${\bf 3}$ &${\bf 3}$ &${\bf 3}~~$\\
${\bf R}$ & $1$&$1$&$0$&$0$&$0$&$0$&$0$&$0$&$0$&$0$&$0$&$0~~$\\ 
${\bf Z}_{2}$ &$+$&$+$&$+$&$+$&$+$&$+$&$-$&$-$&$+$&$-$&$+$&$-~~$\\ 
${\bf U(1)}$ &$0$&$0$&$0$&$-6$&$8$&$2$&$1$&$-4$&$-5$&$-2$&$6$&$5~~$\\ \hline
\end{tabular*}%
\caption{{\footnotesize Transformation of the superfields under the
$SU(3)$ family and $R\times Z_2 \times U(1)$ symmetries which
restrict the form of the mass matrices. 
\label{Table1}}} 
\end{table}%
}
As a second example we will use the model proposed in reference \cite{king2}.
This model is based in a gauged $SU(3) \times SO(10)$ broken to a Pati-Salam
subgroup. The additional symmetries needed to obtain the correct Yukawa textures and Majorana masses  
are given by a $Z_2 \times U(1) \times R$ symmetry
with the charges specified in Table \ref{Table1}. With these symmetries 
the Yukawa superpotential is,
\bea
W_{{\rm Yuk}} &\sim & H \psi_i \psi^c_j \left[ \theta_3^i \theta_3^j \, + \, 
\theta_{23}^i \theta_{23}^j \Sigma \, + \, 
(\epsilon^{i k l} 
\overline{\theta}_{23,k} \overline{\theta}_{3,l}
\theta^j_{23} +\epsilon^{j k l} 
\overline{\theta}_{23,k} \overline{\theta}_{3,l}
\theta^i_{23}) 
\left(\theta_{23} \overline{\theta _{3}}\right) \, +\, \right.  \nn \\
&+&\left. \epsilon ^{ijk}\overline{\theta _{23}}_{k} 
(\theta _{23}\overline{\theta }_{3})^{2} \, +\, 
\epsilon ^{ijk}\overline{\theta _{3}}_{k} 
(\theta _{23}\overline{\theta }_{3})(\theta _{23}\overline{\theta }_{23})
\,+\,\left(\theta _{23}^{i}\theta_{3}^{j}
+\theta _{3}^{i}\theta _{23}^{j}\right) S
\right]
\label{Wking} 
\eea
\bea
W_{{\rm Maj}} &\sim &\Frac{\psi^c_i  \psi^c_j}{M}
\left[\lambda^{i}\lambda^{j}~ +~ 
\theta _{23}^{i}\theta_{23}^{j}(\lambda\overline{\theta
_{23}})(\lambda \overline{\theta_{3}})(\theta _{3}\overline{\theta
_{23}})^{3}~  + \right.\nn \\
&&\left.\frac{1}{M^{13}}
(\epsilon ^{ijk} \overline{\theta _{23,j}}
\overline{\theta _{3,k}})  (\epsilon ^{jlm} \overline{\theta _{23,l}}
\overline{\theta _{3,m}}) (\lambda \overline{\theta _{23}})
(\lambda \overline{\theta _{3}})(\theta _{3}\overline{\theta _{23}})
(\theta _{23}\overline{\theta _{3}})^{2}\right]
\label{Mking}
\eea
Now, if we compare the Yukawa superpotential, Eq.~(\ref{Wking}), with 
the superpotential given in
Eq~(\ref{YukawaWI}) for Model I it is clear that the Yukawa
superpotentials are indeed very similar and the only difference is in
the term that mixes $\theta_3^i \theta_{23}^j$.
Clearly the magnitude is such that it does not affect the quark 
masses and mixing angles much,
although we have seen it can be significant in determining the prediction 
of the phase $\Phi_2$.
On the other hand, the Majorana superpotentials \eq{MajI} and
\eq{Mking} are very different and this modifies the structure of
neutrino masses and mixing angles.

The vev of the $\lambda$ fields break lepton number and gives rise to the Majorana matrix, 
\begin{equation}
M_{RR}\approx \left(\begin{array}{ccc}e^{- i \delta_1/2} & &  \\ & 
e^{- i \delta_2/2} &  \\ & &1\end{array} \right) 
\cdot \left( 
\begin{array}{ccr}
\varepsilon ^{6}\bar{\varepsilon}^{3} & 0 & 0 \\ 
0 & \varepsilon ^{6}\bar{\varepsilon}^{2} & 0 \\ 
0 & 0 & 1%
\end{array}%
\right) \cdot \left(\begin{array}{ccc}e^{- i \delta_1/2} & &  \\ & 
e^{- i \delta_2/2} &  \\  & &1\end{array} \right) M_{3}.  \label{MRRA}
\end{equation}
with $\delta_1 =- 4 \beta_2^\prime - 5 \alpha_u $ and $\delta_2 =
- 4  \beta_2^\prime - \alpha_u + 4\beta_3 $.
Here, the messenger mass scale is the same as in the up sector due to
$SU(4)_{PS}$ and we have factorised two diagonal phase matrices that,
as shown below, play no role in leptonic $CP$ violation. 

Apply the see-saw formula again we obtain the low energy
neutrino mass matrix. As in \cite{king2}, using
the analytic formulas in \cite{King:2002nf} and \cite{King:2002qh},
we obtain
\begin{eqnarray}
m_{1} &\sim &\bar{\varepsilon}^{2}\frac{v_{2}^{2}}{M_{3}} \\
m_{2} &\approx &\frac{(g+\frac{h}{3})^{2}}{s_{12}^{2}}\frac{v_{2}^{2}}{M_{3}}%
\sim 4.8\frac{v_{2}^{2}}{M_{3}} \\
m_{3} &\approx &\frac{[(g-\frac{h}{3})^{2}+(g+\frac{h}{3})^{2}]}{\bar{%
\varepsilon}}\frac{v_{2}^{2}}{M_{3}}\sim 13.5\frac{v_{2}^{2}}{M_{3}}
\end{eqnarray} 
\begin{eqnarray}
\tan \theta _{23}^{\nu } &\approx &\frac{(g+\frac{h}{3})}{(g-\frac{h}{3})}%
\sim 1.2 \\
\tan \theta _{12}^{\nu } &\approx &(\bar{\varepsilon})^{1/2}\frac{(g+\frac{h}{3}%
)}{-cs_{23}}\sim 0.6 \\
\theta _{13}^{\nu } &\approx &-(\bar{\varepsilon})\frac{(g+\frac{h}{3})[(g-%
\frac{h}{3})(-\frac{\alpha }{\varepsilon })+(g+\frac{h}{3})(-\frac{\alpha }{%
\varepsilon }+c\bar{\varepsilon}^{-1/2})]}{[(g-\frac{h}{3})^{2}+(g+\frac{h}{3}%
)^{2}]^{3/2}}\sim 2\bar{\varepsilon}
\end{eqnarray}%
where the numerical estimates correspond to $g\sim 1$, $h\sim 0.3$, $%
c^{\prime }\sim -c\sim 1$, $\alpha \sim \varepsilon \sim 0.05$, 
$\bar{\varepsilon}\sim 0.15$.

To obtain the $CP$ violating phases in the neutrino sector we proceed
as in the previous case. The effective neutrino matrix in the flavour basis 
(before diagonalising the charged lepton Yukawa matrix) is,
\bea
\chi = P_\nu \cdot \left( 
\begin{array}{ccr}
X_1^2 e^{i (\tilde\delta_1 - \delta_2)} & X_1& X_1 \\ 
X_1 & \Frac{X_1^2}{\bar\varepsilon} + e^{i (\delta_2 - \tilde\delta_1)}& 
\Frac{X_1X_2}{\bar\varepsilon} + e^{i (\delta_2 - \delta_1)} \\ 
X_1 & \Frac{X_1X_2}{\bar\varepsilon} + e^{i (\delta_2 - \tilde \delta_1)} & 
\Frac{X_2^2}{\bar\varepsilon} +e^{i (\delta_2 - \tilde \delta_1)}
\end{array}%
\right) \cdot P_\nu 
\eea
with $\tilde \delta_1 = \delta_1 + 2 \delta^u$ and
$P_\nu = \mbox{Diag}(e^{i (\delta^u +\delta_2 - \tilde\delta_1/2)},
e^{i \tilde\delta_1/2},e^{i (\tilde \delta_1/2 + \beta_3)})$. 
Here, we have neglected 
subdominant terms suppresed at least by $\bar \varepsilon^3$.
Now we go to the basis of diagonal charged lepton masses using 
\eq{chargedrotation} and we obtain, 
\bea
\label{chifinmodII}
\chi^\prime = \tilde P  \cdot \left(\begin{array}{ccc}1  &- \bar \varepsilon
\frac{X_1}{\Sigma_e} e^{i \varphi_L}
& 0 \\ \bar \varepsilon \frac{X_1}{\Sigma_e}
e^{-i \varphi_L}&
1 & 0 \\ 0 & 0 &1\end{array} \right) \cdot \chi \cdot 
\left(\begin{array}{ccc}1  &\bar \varepsilon
\frac{X_1}{\Sigma_e} e^{-i\varphi_L} & 0 \\ 
-\bar \varepsilon \frac{X_1}{\Sigma_e}e^{i \varphi_L}&
1 & 0 \\ 0 & 0 &1\end{array} \right) \cdot \tilde P 
\eea
where now $\varphi_L = 2 (\alpha_u - \alpha_d) + 4 \alpha_u + 4 \beta_3$
and $\tilde P = \mbox{Diag}(e^{i(2 \alpha_u -2 \alpha_d +\delta_2 - 
\delta_1/2)}, e^{i \delta_1/2},e^{i(\delta_1/2 -2 \chi + 2\beta_3)})$.
In this basis we calculate the MNS phase through the invariant 
Im$(H^\nu_{12}H^\nu_{23}H^\nu_{13})$. We can use the formula in \eq{invariant}
although, in this case the $A_{ij}$ are no longer real and have a phase at
order $\bar \varepsilon$ with respect to the real part. In this case,
$A_{ij} = \sum_{a=1,3} \chi_{i a} \chi_{j a}^*$ are,
\bea  
A_{12}& =& 2 \frac{X_1^2}{\bar \varepsilon} + X_1 ( 2 + X_1) 
e^{- i (\delta_2 - \tilde \delta_1)} + \dots \nn \\
A_{13}& =& 2 \frac{X_1 X_2}{\bar \varepsilon} + X_1 ( 2 + X_1) 
e^{i (\delta_2 - \tilde \delta_1)} + \dots \nn \\ 
A_{23}&=&\frac{X_1 X_2}{\bar \varepsilon^2} \left(X_1^2 + X_2^2\right) + 
2 \frac{X_2}{\bar \varepsilon} e^{i (\delta_2 - \tilde \delta_1)} +
2 \frac{X_1}{\bar \varepsilon}
e^{- i (\delta_2 - \tilde \delta_1)} + \dots  
\eea
However, we can easily see that $\rm{Im}(A_{12}A_{23} A_{31})$ is 
$O(1/\bar \varepsilon^3)$ while the leading contribution to the invariant is,
\bea
\rm{Im}(H^\nu_{12}H^\nu_{23} H^\nu_{31})\approx 4 \frac{X_1}{\Sigma_e}
\frac{X_1^2 X_2^2}{\bar \varepsilon^4} (X_1^2 + X_2^2) ( X_1^4 - X_2^4
+ 2 X_1^2 X_2^2) \sin \varphi_L + \dots
\label{MSNfinII}
\eea
and the real part of the invariant is given at leading order by
$4 X_1^4 X_2^2 (X_1^2 + X_2^2)/\bar \varepsilon^4$. 
Therefore in this case the MNS phase is given by $\varphi_L = 2 \alpha_d 
- 2 \alpha_u -2 \beta_3 + 2 \beta^\prime_2$ and again charged 
lepton phases play a dominant role in the MNS phase 

Now, we must calculate the leptogenesis phase from \eq{leptogen}
in the basis of diagonal Majorana masses, \eq{dirac'}. Again in this case, 
the phases coming from charged leptons cancel and therefore this phase is 
unrelated to the MNS phase.  

Finally the neutrinoless double beta decay phase is the relative phase between the two dominant contributions to $\chi'_{11}$. We have
\bea
(\chi^\prime)_{11} = \tilde P_1^2 \left(\chi_{11} - 2 \bar \varepsilon 
\frac{X_1}{\Sigma_e} \chi_{12} e^{-i \varphi_L} \right)
\simeq \tilde P_1^2 X_1^2 e^{i (\tilde \delta_1 - \delta_2)} \left( 1 - 
\frac{2\bar \varepsilon}{\Sigma_e} e^{i (\delta_2 - \tilde \delta_1 - \varphi_L)} \right)
\eea
Therefore, the neutrinoless double beta decay phase is simply,
$(\delta_2 - \tilde \delta_1 - \varphi_L) = 2 (\alpha_u - \alpha_d)$. 
This is just the CKM phase in the quark sector.

In summary, the phases are constrained in specific models although the
constraints are model dependent. In Model I the MNS phase, measureable
in neutrino oscillations, is the same as the phase relevant to double
beta decay processes.  However there is no relation with the quark CKM
phase in this case. In Model II the double beta decay phase is just
the CKM phase but now is not related to the MNS phase. In both cases
the leptogenesis phase is not given in terms of the other phases.  As
we shall discuss in Section \ref{sec:vac}, all the phases are
determined at the stage of spontaneous breaking. This leads to a
discrete set of possible values for the phases. For example in Model I
the CKM phase is quantised in units of $\pi/2$ showing that maximal CP
violation in the quark sector is quite natural. All the other phases
are similarly quantised but the unit of quantisation is too small to
make phenomenologically interesting predictions.

\section{Supersymmetric soft breaking and the flavour and $CP$
problems} 
\label{sec:SUSY} 
So far we have seen that an $SU(3)$ model of flavour with spontaneous 
$CP$ violation can successfully reproduce the quark and lepton masses
and mixing angles together with the observed $CP$ violation. 
We have also calculated the predictions for CP violation in the neutrino sector. 
In this Section we analyse the new effects of these phases in the 
supersymmetric sector.
\subsection{The SUSY CP problem}
Perhaps the most severe problem in supersymmetry phenomenology is the
so-called ``supersymmetric $CP$ problem''. This refers to the  
SUSY contributions to the electron and neutron electric dipole moments
(EDM) that in the presence of phases in the $\mu$ term or the trilinear
couplings typically exceed the experimental bounds by two orders of 
magnitude. Therefore, first we must check these phases.

The $\mu$ term in our model is generated through a Giudice-Masiero
mechanism. This means that there is no bilinear in the Higgs fields in the
original superpotential and this term is only present in the K\"ahler
potential as $\lambda H_1 H_2 + h.c.$. 
Once supersymmetry is broken 
an effective $\mu$ term is generated in the superpotential of the form
$\lambda m_{3/2} H_1 H_2$. In our model, $CP$ is an exact symmetry at
$M_{Plank}$, before the breaking of the flavour symmetry. Thus, at 
this scale, the term $\mu_0 = \lambda m_{3/2}$ is real 
\cite{spontCPflavour}. 
Now we have
to worry about the corrections proportional to flavon vevs once we 
break the flavour symmetry. However, it is easy to see that these
corrections will be extremely small and irrelevant for phenomenology.
First we must take into account that in the absence of the $\mu$ term
(or equivalently of the mixed Higgs term in the K\"ahler potential)
our theory has a new U(1) symmetry distinguishing the two Higgs
doublets. Therefore any correction to the $\mu$ term must contain
the $\mu$ term itself. This implies that the first correction to the 
$\mu$ term must come from a two loop diagram \cite{barr}. 
Then only a neutral combination of flavon fields
under both $SU(3)$ and the other flavour symmetries (different from the trivial
ones $\theta_a \theta_a^\dagger$) can contribute and
we have seen in the previous Section that the size of this neutral     
combination is constrained from the structure of the Majorana mass matrix
and can be at most $\varepsilon^{2.75}$.
In summary, the phase of the $\mu$ term is well below $10^{-4}$.

Regarding the trilinear couplings, we can obtain the trilinear couplings
after the flavour symmetry breaking in terms of the effective superpotential
\eq{YukawaW0} and the effective K\"ahler potential \eq{kahler}. Here,
it is much more convenient to work in the basis of diagonal
K\"ahler metric for the visible fields. In this basis, the
trilinear couplings, $(Y^A_\beta)_{ij} H_\beta Q_i q^c_j$, 
are,
\bea
(Y^A_{\beta})_{ij} = Y^{\beta}_{ij} F^a \partial_a \left( \tilde K + \log (
K^{\beta}_\beta K^i_i K^j_j) \right) + F^a \partial_a  \hat Y^\beta_{ij}
\label{trilinear}
\eea
where $Y_{ij}^\beta$ is the Yukawa matrix in terms of the canonically 
normalised fields, $\hat Y^\beta_{ij}$ are the original Yukawa couplings 
before cannonical normalisation and the K\"ahler potential is 
$K = \tilde K (X,X^\dagger) + K^i_i (X, X^\dagger) |\phi^i|^2$ 
with $X$ hidden sector fields,
$\phi$ visible sector fields and $K^i_i (X,X^\dagger)$ real. 

The first term in \eq{trilinear} gives rise to a factorisable contribution
to the trilinear couplings $Y^A_{ij} = (A_i + A_j) Y_{ij}$ \cite{tatsuo} with 
$A_i =F^a \partial_a \log K^i_i$ and $A_j =F^a \partial_a (\tilde K +
\log (K^j_j K^\beta_\beta))$ real. From \cite{khalil}, and taking into account 
that these $A_i$ are real and the only phases are in the Yukawa matrix, 
we can see that the contribution from these factorisable nonuniversal 
trilinear terms
to EDMs is safely below the experimental bounds. The remaining term in
\eq{trilinear} involves the derivative of the Yukawa couplings in terms of
fields with nonvanishing F-terms. If we make the derivative in terms of
the flavon fields themselves it was shown in \cite{flat} that the diagonal
trilinear couplings in the SCKM basis are real at leading order in the
flavon fields. In the same spirit, we can prove that this is also true 
for derivatives in terms of real hidden sector fields. 
If the Yukawa matrix is 
$Y(\theta,X) = V_L(\theta,X)\cdot D(\theta,X)\cdot V_R(\theta,X)$ this
contribution to the trilinear couplngs in the SCKM basis is,
\bea
F^a (V^\dagger_L\cdot \partial_a Y \cdot V_R)_{ii} = 
F^a(V^\dagger_L\cdot \partial_a V_L)_{ii} D_i + F^a \partial_a D_i + 
F^a D_i (\partial V^\dagger_R\cdot V_R)_{ii}
\label{prooftri}  
\eea
Now a given eigenvalue or an element of the mixing matrix can also be 
expanded in terms of the small flavon fields, 
$V_{ij} = \sum_n (a_n + b_n X) \theta^n$.
If we expand \eq{prooftri} in term of $\theta$, the order in $\theta$ of the
eigenvalue will set the leading term of the expansion as it appears in 
the three terms in \eq{prooftri}.  In fact, if $D_i = (a_m + b_m X)~ \theta^m$
for some $m$,
$\partial_a D_i$ will keep the same order in $\theta$, $\propto \theta^m$. 
Moreover, as 
the only phases are in the flavon fields, $\partial_a D_i$ has the same phase 
as $D_i$. Then the first or third terms in this equation, will give a 
leading order contribution to the expansion only if 
$(V^\dagger\cdot \partial_a V)$ does not depend on $\theta$ and and in this 
case is completely real. So, all the leading order contributions keep the 
same phase as the eigenvalue itself and therefore is rephased away when we
make the mass real. Only subdominant
terms in the $\theta$ expansion can get any phase, exactly as in \cite{flat}.
This suppression is already enough to satisfy the EDM bounds. 

\subsection{The SUSY flavour problem}
We have shown that the SUSY $CP$ problem is elegantly
solved in this theory. However, we must also check the status
of the SUSY flavour problem (including offdiagonal $CP$ violation). 

In this case we have two possible sources of new flavour structures
associated with the breaking of the flavour symmetry, either D-term or
F-term contributions. First, we must consider the possibility of large 
D-term contributions that can spoil the required degeneracy among squarks.  
In our models, the various D-terms associated to $SU(3)_{fl}$ breaking are 
small due to the symmetry of the flavon superpotential in the fields 
$\theta_3$--$\overline{theta}_3$ and 
$\theta_{23}$--$\overline{\theta}_{23}$ \cite{king2} as shown in section 
\ref{sec:vac}. So we can neglect these D-term contributions to the 
sfermion masses.
On other hand we also have contributions to soft masses due to some 
nonvanishing F-terms, as we have already discussed in Section 3, \eq{soft1}. 

Thus, we obtain the sfermion mass matrices from the K\"ahler potential
\eq{kahler}, as $M^2_{ij} = m_{3/2}^2~ \delta_{ij} + F_a F^b 
(\partial \log K_{i j})/(\partial \phi_a \partial \phi^b)$. The general 
structure of these matrices is given by \eq{soft1} although the expansion 
parameter is different for different $SU(2)_R$ quantum numbers 
$\bar \varepsilon \simeq 0.15$ for the right handed 
down quarks and charged leptons and $\varepsilon \simeq 0.05$ for 
right handed up quarks as well as all $SU(2)_L$ doublets (for a Pati-Salam
group below $M_{Plank}$). This means that the largest effects will be
present in the right-handed down squark mass matrix and the right-handed
charged slepton mass matrix, although differences in the left handed mass 
matrices can be also significant.  
At this point, we should emphasize that measurement of the complete flavour 
structure of the sfermion mass matrices will be the final check of the nature
of the flavour symmetry. Ideally we would like to produce directly the 
different sfermions and simply measure their masses and mixing angles.
However, before we are able to do that at LHC or at the Next Linear Collider
we can still look for indirect signs of SUSY flavour in FCNC or $CP$ 
violation experiments \cite{annrev}.
    
Using Eqs.~(\ref{thematrix}) and (\ref{soft1}) we can obtain the 
right-handed down squark mass matrix in the SCKM basis where we can
use the phenomenological Mass Insertion bounds. However, this is not 
enough in the case of complex contributions, as we have still the 
freedom to rephase simultaneously left-handed and right-handed 
quarks without changing the eigenvalues. This rephasing does change 
the phase in different elements of the CKM mixing matrix and in fact
in the presence of non universality new Yukawa phases become observable
\cite{piai}. The MI bounds
calculated in \cite{MI} are only valid in a basis where the phases in
the CKM matrix are located in the $(1,3)$ and $(3,1)$
elements (up to order $\bar \varepsilon^4$) \cite{oleg}. 
Therefore we must go to this particular basis to
compare our predictions with the bounds. More explicitly, the down 
Yukawa matrix in \eq{thematrix} is diagonalised, $Y^d = V_L^d \cdot 
Y^d_{diag} \cdot V_R^{d\,\dagger}$, where,
\bea
V_R^d = \left(\begin{array}{ccc}(1-\frac{\bar \varepsilon^2}{2})e^{-i \delta^d} 
&\bar \varepsilon\frac{X_1}{\Sigma_d}e^{-i \delta^d}&
\bar \varepsilon^3X_2e^{-i(\beta - \chi+\delta^d)}\\
-\bar \varepsilon\frac{X_1}{\Sigma_d}
&1-\frac{\bar \varepsilon^2}{2}&\bar \varepsilon^2\Sigma_d e^{-i(\beta - \chi)}\\
\bar \varepsilon^3 2 X e^{-i(2 \chi -\beta)}&-\bar \varepsilon^2\Sigma_d 
e^{-i(2\chi - \beta)}&e^{-i \chi}\end{array} \right) \cdot
\left(\begin{array}{ccc}e^{i \omega }&0&0 \\0&1&0\\
0&0&e^{-i (\beta - \chi)}\end{array} \right)
\label{SCKMrephase}
\eea
and 
\bea
V_L^d = \left(\begin{array}{ccc}(1-\frac{\bar \varepsilon^2}{2})e^{i \delta^d} 
&\bar \varepsilon\frac{X_1}{\Sigma_d}e^{i \delta^d}&
\bar \varepsilon^3X_2e^{i(\beta - \chi+\delta^d)}\\
-\bar \varepsilon\frac{X_1}{\Sigma_d}
&1-\frac{\bar \varepsilon^2}{2}&\bar \varepsilon^2\Sigma_d e^{i(\beta - \chi)}\\
\bar \varepsilon^3 2 X e^{i(2 \chi -\beta)}&-\bar \varepsilon^2\Sigma_d 
e^{i(2\chi - \beta)}&e^{i \chi}\end{array} \right) \cdot
\left(\begin{array}{ccc}e^{i \omega }&0&0 \\0&1&0\\
0&0&e^{-i (\beta - \chi)}\end{array} \right)
\label{SCKMLrephase}
\eea
where the diagonal phase matrices clearly do not affect the Yukawa
matrix in  $Y^d = V_L^d \cdot Y^d_{diag} \cdot V_R^{d\,\dagger}$, but
combined with different rephasings in the up sector bring the
CKM mixing matrix to the standard form with phases in $(1,3)$ and $(3,1)$
elements and $\omega = \mbox{Arg}\left(1 + 
\bar \varepsilon \frac{\Sigma_d}{\Sigma_u} e^{i \Phi_1}\right)$.

Now we have $(M^2_{\tilde{D}_R})^{SCKM} = V_R^{d\,\dagger}\cdot 
M^2_{\tilde{D}_R} \cdot V_R^d $
\bea
(M^2_{\tilde{D}_R})^{SCKM} \simeq \left(
\begin{array}{ccc} 1 + \bar \varepsilon^3 & - \bar \varepsilon^3 
e^{-i \omega}\Frac{|a^d_3|^2}{\Sigma_d} X_1 &- \bar \varepsilon^3 
e^{-i \omega}\Frac{|a^d_3|^2}{\Sigma_d} X_1  \\ 
- \bar \varepsilon^3 e^{i \omega}\Frac{|a^d_3|^2}{\Sigma_d} X_1 & 
1 + \bar \varepsilon^2 & \bar \varepsilon^2 - \bar \varepsilon^3 \Sigma_d 
e^{- i 2 (\beta - \chi) } \\
- \bar \varepsilon^3 e^{i\omega }\Frac{|a^d_3|^2}{\Sigma_d} X_1  
& \bar \varepsilon^2 - \bar \varepsilon^3 \Sigma_d e^{ i 2 (\beta - \chi) }& 
1 + \bar \varepsilon \end{array}
\right)  m_0^2
\label{softm2}
\eea
In this case, we obtain a large $(\delta_{RR})_{12}$ from the rotation of the
``large'' difference among $(1,1)$ and $(2,2)$ elements, that
generates the order $\bar \varepsilon^3$ entry in the $(1,2)$ element.
The modulo of this entry is (for $\bar \varepsilon = 0.15$ and $X_1=1.4$) 
is $5 \times 10^{-3}$ at $M_{GUT}$.  For the imaginary part we obtain 
\bea
\mbox{Im} (M^2_{\tilde{D}_R})_{12} = \bar \varepsilon^4 \frac{|a^d_3|^2}{\Sigma_u} 
X_1 \sin \Phi_1 m_0^2 \simeq 1.15 \times 10^{-3} m_0^2
\eea
where we taken $\Phi_1 = 90^0$ and $|a^d_3|^2/\Sigma_u\simeq 1.5$.
We must take into account that off diagonal elements of squark
mass matrices do not change much from $M_{GUT}$ to $M_W$. However 
the diagonal elements get a large gaugino contribution and then, 
at the electroweak scale we can roughly approximate the average 
squark mass by $5 m_0^2$. So we finally estimate the value of the
Mass Insertion at the electroweak scale as, $\mbox{Re} (\delta^d_{RR})_{12} 
\simeq 1 \times 10^{-3}$ and $\mbox{Im} (\delta^d_{RR})_{12} 
\simeq 2.3 \times 10^{-4}$.

This must be compared with the phenomenological MI bounds \cite{MI}. 
In this case, the relevant bound is,
\bea
\left(\sqrt{|\mbox{Im} (\delta_{RR}^d)^2_{12}|}\right)^{model} = 6.8 
\times 10^{-4} \leq
\left(\sqrt{|\mbox{Im} (\delta_{RR}^d)^2_{12}|}\right)^{bound} = 3.2 
\times 10^{-3}  
\eea

In a similar way, we can obtain the left handed down squark mass matrix
in the SCKM basis. We must start from this matrix in the flavour symmetry
basis, that now has a smaller expansion parameter,
$\varepsilon\simeq 0.05$.
\bea
\label{softL}
M^2_{\tilde{D}_L} \simeq \left(
\begin{array}{ccc} 1 + \varepsilon^3 & 0 & 0 \\ 
0 & 1 + \varepsilon^2 & \varepsilon^2 e^{i \beta_3} \\
0 & \varepsilon^2 e^{-i \beta_3} & 1 + \varepsilon \end{array}
\right)  m_0^2
\eea
and then with the rotation \eq{SCKMLrephase} we obtain in the SCKM
basis, $(M^2_{\tilde{D}_L})^{SCKM} = V_L^{d\,\dagger}\cdot 
M^2_{\tilde{D}_L} \cdot V_L^d$,
\bea
(M^2_{\tilde{D}_L})^{SCKM} \simeq \left(
\begin{array}{ccc} 1 + \varepsilon^3 & - \bar \varepsilon \varepsilon^2
e^{-i \omega}\Frac{|a^d_3|^2}{\Sigma_d} X_1 &- \bar \varepsilon \varepsilon^2 
e^{i(2 \chi- \omega)}\Frac{|a^d_3|^2}{\Sigma_d} X_1  \\ 
- \bar \varepsilon \varepsilon^2 e^{i \omega}\Frac{|a^d_3|^2}{\Sigma_d} X_1 & 
1 + \varepsilon^2 & \varepsilon^2 e^{i 2 \chi} - \varepsilon 
\bar \varepsilon \Sigma_d \\
- \bar \varepsilon \varepsilon^2 e^{-i( 2 \chi-\omega)}
\Frac{|a^d_3|^2}{\Sigma_d} X_1  
& \varepsilon^2 e^{i 2 \chi} - \varepsilon 
\bar \varepsilon \Sigma_d & 1 + \varepsilon \end{array}
\right)  m_0^2
\label{softm2Lbis}
\eea

The presence of offdiagonal entries in these squark mass matrices
can have sizeable effects in several low energy observables.
We have already analysed the effects of RR mass insertion in 
$\varepsilon_K$ and we have seen they can reach a $20\%$. However,
if we consider simultaneously the RR and LL mass insertions the 
contribution could be even larger. For this combination the 
phenomenological bound is, 
$\sqrt{|\mbox{Im} (\delta_{RR}^d)_{12} (\delta_{LL}^d)_{12}|}
\leq 2.2 \times 10^{-4}$, while in our case even with the smaller
LL mass insertion we obtain at the electroweak scale, 
\bea
\sqrt{|\mbox{Im} (\delta_{RR}^d)_{12} (\delta_{LL}^d)_{12}|} =
\sqrt{ \mbox{Im}\left( ( 1 + i \, 0.23) 10^{-3}\ ( 1 + i\,0.23) 10^{-4}
\right)} = 2 \times 10^{-4}
\eea 
for $\Phi_1 = 90^0$. This can give a large part of the experimental
result. However, we must take into account that this represents the
maximum possible contributions and all the elements in the sfermion mass 
matrices are multiplied by unknown factors order 1. Nevertheless we
can  conclude that a sizeable contribution to $\epsilon_K$ can be expected.

In the case of $\delta_{13}$ MI that could contribute to $B_d \bar B_d$ 
mixing and the $J/\psi K_S$ $CP$ asymmetry, we can see from \eq{softm2} 
and \eq{softm2Lbis} that these MI are exactly of the same order as the 
corresponding $\delta_{12}$ MI. However, the values for the MI required
to saturate these observables are now much larger and the phenomenological
bounds are, $\sqrt{|\mbox{Im} (\delta_{RR}^d)^2_{13}|} \leq 0.3$  and 
$\sqrt{|\mbox{Im} (\delta_{RR}^d)_{13} (\delta_{LL}^d)_{12}|} \leq 9 
\times 10^{-3}$ \cite{reyes}. Thus no sizeable effects are possible here. 

Finally we can consider $b \to s$ transitions with a $\delta_{23}$, 
contributing to $B_s \bar B_s$ mixing and some $CP$ asymmetries as
$B \to \Phi K_S$ \cite{phiks}. From Eqs.~(\ref{softm2}) and
(\ref{softm2Lbis}) we obtain
 at the electroweak scale,
\bea
(\delta^d_{RR})_{23} \simeq (4.5 + 6.8 \, i)  \times 10^{-3} ~~~~
(\delta^d_{LL})_{23} \simeq (1.5 + 0.5 \, i)  \times 10^{-3}
\eea
where, once more, we assumed maximal phases. Although in this case
there are no MI bounds as in \cite{MI,reyes}, from the literature
on SUSY contributions to this decay \cite{bphiks, ciuchini}, it is clear that
an isolated LL or RR mass insertions needs to be at least 
${\cal{O}}(0.2)$ to have a sizeable effect. It is not clear in the case
of simultaneous LL and RR MI, but from \cite{ciuchini} it seems that
a MI of order 0.05 is required. Therefore, apparently there is no sizeable
effect here.

Lepton flavour violating decays, as $\mu \to e \gamma$ and 
$\tau \to \mu \gamma$, can also receive sizeable contributions.
In this case, the slepton mass matrices are exactly identical to
Eqs.~(\ref{softm2}) and (\ref{softm2Lbis}), with the only replacement
of $\Sigma_d \to \Sigma_e$ (in this case the presence of phases is not 
important). However, the main advantage of leptonic processes is that
the MI are not largely reduced from $M_{GUT}$ to $M_W$, 
as the diagonal elements of slepton mass matrices are less afected by
gaugino  masses in the evolution between these two scales, in fact at
$M_W$ we can approximate the average left handed slepton mass by
$2\, m_0^2$. In this case, the bounds on the leptonic MI are more 
difficult to obtain as they depend on other parameters ($\mu, \tan \beta, 
M_{1/2}, m_0^2)$. However, there some bounds for fixed values of
$\tan \beta$ \cite{LeptMI}. The most sensitive precess is $\mu \to e \gamma$ 
where for $\tan \beta =10$ and average slepton mass of $300 GeV$ we get
$(\delta^e_{LL})_{12} \leq 3 \times 10^{-4}$ while for the RR mass insertion
the bound is much worse due to a possible cancelation among diagrams. 
Now we have for the LL MI from \eq{softm2Lbis}, 
$(\delta^e_{LL})_{12} \leq 2.5 \times 10^{-4}$ and therefore a 
$\mu \to e \gamma$ decay close to the experimental bound is indeed possible.

Finally, we have to comment on possible nonuniversality effects in the 
trilinear couplings. Trilinear couplings are obtained from \eq{trilinear}
and in this case, it is easy to see \cite{KKV,KvsB} that large effects 
can be expected again in the kaon and leptonic sectors due to the big
experimental sensitivity in these experiments. However these effects are 
usually sufficiently small in the B system due to the associated suppression 
with quark masses and the smaller sensitivity in B experiments. 
Again, as shown in \cite{flat},
these effects can be close to the experimental bound in the decay 
$\mu \to e \gamma$ if the flavon F-terms are sizeable.   

\section{Vacuum Alignment}

\label{sec:vac}

It is easy to construct a superpotential capable of generating the phenomenologically desirable pattern of symmetry breaking vevs. 
Since in \cite{king2}  a suitable superpotential for Model II has been constructed, we concentrate here on constructing a suitable superpotential for Model I. We will use it to demonstrate how maximal CP violation may readily be obtained. 

There is considerable freedom in constructing the breaking sector because it is sensitive to additional family and SM gauge singlet fields. Such fields abound in compactified string theories so we make no apologies for adding several such fields to the theory. An example of a set (we have not searched for the minimal set) giving rise to acceptable spontaneous breakdown is given in  Table \ref{tab:higgs2}. The resulting superpotential has the form
\begin{table}[tbp] 
\begin{tabular*}{1.0\textwidth}{@{\extracolsep{\fill}}|c| c c c c c c c c 
c|}
\hline
~${\bf Field}$  &$P$ &$S$ &$\overline{S}$&$ T$ & $U$& $V$ &$Y$&$Z$ & $\overline{V}$~~~\\ 
~${\bf SU(3)}$ & $1$& $1$& $1$& $1$& $1$ & $1$& $1$& $1$&$1$~~~ \\
~${\bf U(1)}$ & $-2$ & $-9/2$ &  $9/2$ &  $0$ & $7$ & $-2$ &$-11$& $-11$
&$-11$~~~ \\
~${\bf Z_{5}}$& $3$& $2$ & $1$& $2$ & $1$  & $2$ &$4$ &$2$&$1$ ~~~\\
~${\bf Z_{3}}$& $1$& $2$ & $0$& $2$ & $2$  & $1$ &$1$ &$0$&$1$ ~~~\\
\hline
\end{tabular*}
\caption{Transformation of the superfields associated with the 
spontaneous symmetry breaking.}
\label{tab:higgs2}
\end{table}
\bea
W &=& P(\theta_{3} \bar \theta_{3}+T)+ U \left( (\theta_{23} \bar \theta_{23}) +  S^2\right) \, + \, 
          V \left((\theta_{3} \bar \theta_{3})^4 +  S \bar S \right) \, + \,
            \nn \\
&& \, + \,Y (\theta_{3} \bar \theta_{2}) \, + \, \bar V \bar S ^2 
(\theta_{3} \bar \theta_{3})^4  \, + \, 
Z \left( (\theta_{23} \bar \theta_{3})(\theta_{23} \bar \theta_{2}) + 
\bar S^2 \right)  \nn \\
&& \, + \,  \mu H_1 H_2 \left[ 1 \, +
\,(\theta_{23} \bar \theta_{3}) (\theta_{3} \bar \theta_{3})^3 \, + \, T^5+...
 \right]
\label{higgsW}
\eea
The overall symmetry breaking here is triggered by the fields $\theta_3$, 
$\bar\theta_3$ which obtain a vevs along a flat direction through 
radiatively breaking due to the Yukawa coupling $P\theta_{3} \bar \theta_{3}$ 
in \eq{higgsW}.  Then it is transmitted to the rest of the
flavon sector through the other terms in the superpotential. Note that 
the fields $\theta_3$, $\bar\theta_3$ couple symmetrically so that, if 
their soft masses are degenerate at the Planck scale, they will remain 
degenerate. As we show below, this is important in ensuring the close 
equality of their vevs which is necessary in avoiding large family 
dependent D-terms and the associated flavour changing processes \cite{king2}.
Notice however, that a possible non-degeneracy of the soft masses would only 
change the relative size of the vevs in factors ${\cal{O}}(1)$ but 
never their order in $\varepsilon$ and our predicted Yukawa textures 
would not be affected.
 
From this flavon superpotential, we have the following scalar potential,
\bea
V = \left|F_U\right|^2\, + \,\left|F_V\right|^2\, + \,
\left|F_{\bar V}\right|^2\, + \, \left|F_P\right|^2\, + 
\,\left|F_Y\right|^2\, + \,\left|F_Z\right|^2 \nn \\
 \, + \, m_{3/2}~ \mu~ H_1 H_2\, \left[ 1 \, +\, 
(\theta_{23} \bar \theta_{3}) (\theta_{3} \bar \theta_{3})^3 \, + \, T^5
 \right]
\label{flavonpot}
\eea
where the last line corresponds to the soft terms associated with
\eq{higgsW}. Note that we have included in Eqs.~(\ref{higgsW}) and 
(\ref{flavonpot}) a $\mu$ term which only appears after SUSY breaking
through a Giudice-Masiero mechanism. Notice that although this contribution 
to $V$ is extremely small
it is the first contribution lifting the flat direction left by the F-terms. 
Clearly, any neutral combination of
vevs can multiply the superpotential but the terms in \eq{higgsW} are
enough to specify all the relevant phases in this case.

After the fields $\theta_3$, $\bar\theta_3$ get radiatively induced vevs, 
minimisation of $F_V$ generates vevs for $S$ and $\bar S$ which are equal, 
$\langle S \rangle =\langle \bar S \rangle$, if they have
equal soft breaking masses.  Then minimisation of $F_{\bar V}$ forces the 
alignment $(\theta_{3}^u 
\bar \theta_{3}^u)/M_{u}^2 = (\theta_{3}^d \bar \theta_{3}^d)/M_{d}^2$ 
and minimisation of $F_U$ generates 
a vev for $(\theta_{23} \bar \theta_{23})$. 
Using this together with D-flatness, we obtain,
\bea
\Frac{|\langle \theta_{23} \rangle |}{M_{u}} = \Frac{|\langle \bar 
\theta_{23} \rangle |}{M_{u}} = \varepsilon, ~~~~~~~~~
\Frac{|\langle \theta_3^{u} \rangle |}{M_{u}} = \Frac{|\langle \bar 
\theta_3^{u} \rangle |}{M_{u}} = \varepsilon^{1/4} \\
\Frac{|\langle \theta_{23} \rangle |}{M_{d}} = \Frac{|\langle \bar 
\theta_{23} \rangle |}{M_{d}} = \bar\varepsilon, ~~~~~~~~~
\Frac{|\langle \theta_3^{d} \rangle |}{M_{d}} = \Frac{|\langle \bar 
\theta_3^{d} \rangle |}{M_{d}} = \varepsilon^{1/4} \nn
\eea
Minimisation of $F_Z$ forces $\bar \theta_2$ to acquire a vev while
minimisation of $F_Y$ makes it orthogonal to $\theta_3$, by definition
the 2 direction. Thus we see that minimisation of $F_Z$ forces
$\theta_{23}$ to get vevs in the 2 and 3 directions. Finally,
inclusion in the scalar potential of the SUSY breaking soft mass terms
for $\theta_{23}$ (due to the $SU(3)$ family symmetry they are equal
in the 2 and 3 directions) shows that $\theta_{23}$ will get {\it
equal} vevs in the 2 and 3 directions - the required vacuum alignment.

A major problem in models with a continuous gauged family symmetry is
that the associated D-terms could split the degeneracy among different
sfermions. Consider, for instance, the fields  $\theta_{23}$ and 
$\bar \theta_{23}$. Making $F_U$ zero
fixes the vev of the product $(\theta_{23} \bar \theta_{23})$. Then, if the 
soft masses of these fields are equal we can also cancel the D-terms
choosing equal vevs for the barred and unbarred fields. If the soft masses 
are not equal the size of the D-terms are roughly determined by the 
difference of the soft masses.  Therefore, to 
minimise the value of the D-terms, we must assure that the soft masses 
of the relevant fields are reasonably close at the scale of symmetry breaking. 
These soft masses must be approximately equal at $M_{Plank}$ and this 
degeneracy must not be spoiled by radiative corrections between $M_{Plank}$ 
and the symmetry breaking scale. Regarding the required level of degeneracy, 
$D_{23}$ is precisely the most dangerous D-term,
\bea
|D_{23}|^2 =  g^2 | \theta_{23}^2 \theta_{23}^{3\,*} - \bar \theta_{23,2} 
\bar \theta_{23,3}^* + \psi_2 \psi_3^\dagger |^2,
\eea
as this would give an unsuppressed off-diagonal contribution to the squark 
masses. However, as seen in the previous section, the phenomenological
bounds on this off-diagonal entry are not too strong. Moreover due to large
gaugino contributions in the running from $M_{Plank}$ to $M_W$ to diagonal
squark masses, the relative size of the off-diagonal entries is reduced.
Therefore, we only have to require a mild degeneracy ${\cal{O}}(0.1)$
to the $\theta_{23}$ and $\bar \theta_{23}$ soft masses \footnote{This 
D-term would also generate an offdiagonal entry in the $(1,2)$ element 
of the squark mass matrix after diagonalisation of the corresponding 
Yukawa matrix. However, the phenomenological constraints are satisfied 
with the same level of degeneracy.}. This level of 
degeneracy is quite natural in string theory where groups of fields are 
expected to be degenerate. Another 
possibility that can also readily occur in string theory is that
the soft masses of  $\theta_{23}$ and $\bar \theta_{23}$ are  
$\lsim {\cal{O}}(0.1)~ m_{\tilde q}^2$, with $m_{\tilde q}$ the soft mass 
associated with the squark fields. Even if we assume that these masses
are exactly equal at $M_{Plank}$ the degeneracy will be broken by radiative 
corrections if these fields have different interactions. 
However, as we have noted above, the fact that that the
superpotential in \eq{higgsW} is approximately symmetric in $\theta_3$
and $\bar \theta_3$ or $\theta_{23}$ and $\bar \theta_{23}$ implies
their soft masses will not receive significantly different radiative
corrections from $M_{Plank}$ to the symmetry breaking scale. As a result 
\cite{king2} the $SU(3)_{fl}$ D-terms are small and consistent with the 
bounds from FCNCs.

Finally, following the discussion of \cite{king2}, the lepton number
violating vevs of $\lambda$ and $\bar\lambda$ are automatically
aligned along the 3 direction.

We turn now to the really new feature, namely the spontaneous
determination of the phases. They are also fixed from the minimization
of the scalar potential. We first remind the reader that, due to the
underlying CP symmetry, the suppressed couplings in
Eq.~(\ref{flavonpot}) are real. Thus the minimisation of F terms
involving the cancellation of two contributions requires that the
relative phases of the two terms must be $\pm 1$. Then from the
minimisation of the F-terms in Eq.~(\ref{flavonpot}) we obtain
\bea
\theta_T &\approx& \alpha_u + n_1~\pi \nn \\
\theta_S &=&  \beta^\prime_2 + n_2 ~\pi  \\
\theta_S\,+\, \theta_{\bar S} &=&  4 ~\alpha_u + n_3~ \pi \nn \\
4(\alpha_u - \alpha_d - \chi) &=&n_4 ~\pi \nn \\
2~ \theta_{\bar S} &=&   \alpha_u \,+\, \beta_3 \,+\, \delta_2 + n_5~ \pi
\nn
\eea
Further from the soft terms in the scalar potential,
\bea
\label{phtri}
\beta_3+4\alpha_u=m_1~\pi \nn \\
3( \alpha_u \,-\,~ \alpha_d-\chi)+(\alpha_u \,-\,~ \alpha_d) &=&  m_2 ~\pi  \nn \\
5~ \theta_T &=& m_3 ~\pi 
\eea
with $n_i$, $m_i$ integers. 

From this system of equations we obtain,
\bea
\theta_T~ =~ m_3 \frac{\pi}{5},~~~~~ 
\alpha_u~ -~ \alpha_d~ = (m_2-\frac{3}{4}n_4)\pi,~~~~~~
\eea
The CP violating phase in the CKM matrix is 
$\Phi_1=2(\alpha_u-\alpha_d)=-\frac{3}{2}m_4\pi$. 
We see that it is quantised in units of $\pi/2$ and so it is quite 
natural to obtain maximal 
CP violation in the quark sector in good agreement with observation.

One may proceed in this manner to determine the other phases. Unfortunately, 
although all the other phases are also quantised,
the unit of quantisation is too small to make phenomenologically 
interesting predictions.

\section{Conclusions}

In this paper we have investigated spontaneous $CP$ violation in models based on an
$SO(10)\times SU(3)_{fl}$ symmetry. We have shown that such models provide 
a remarkably consistent description of the known masses and mixing angles of quarks 
and leptons. In the quark sector, the presence of $CP$ violating phases
is necessary, not only to reproduce $CP$ violation processes, but
also to reproduce the observed masses and mixings.
We have also shown that the spontaneous breaking of $CP$ in the flavour sector
naturally solves the supersymmetric $CP$ problem and the SUSY flavour
problem, although flavour changing processes must occur at a level close to current experimental bounds.

We have presented two possible variations of the model which simultaneously reproduce the
observed neutrino masses and mixings. In these models there are phenomenologically interesting
relations between the CP violating phases in the quark and lepton sectors and testing these relations will
be a sensitive discriminator of models. The magnitude of CP violation is determined by the potential 
driving the spontaneous family symmetry breaking. The structure of this potential is such that it can quite naturally 
lead to maximal violation in the quark sector, consistent with present observations.
An important complication, that we expect to be quite generally true, is that charged lepton phases, even if
there is relatively small charged lepton mixing, cannot be ignored
in the determination of the MNS and neutrinoless double beta decay phases. On the other hand the
leptogenesis phase is independent of the charged lepton phases.

The main prediction of the models based on a non-Abelian family group is the 
full flavour and $CP$ structure of the sfermion mass matrices. 
Accurate measurement of sfermion masses and 
mixing angles will provide a significant test of such models. Even before we are able to access the sfermion sector 
at the LHC or at the Next Linear Collider there are several FCNC or $CP$
violation experiments receive which sizeable SUSY contributions in such models and which already provide an experimental probe
of the underlying family symmetry.

\section*{Acknowledgements}
This work was partly funded by the PPARC rolling grant PPA/G/O/2002/00479 and 
the EU network \textquotedblleft Physics Across the Present Energy Frontier", 
HPRV-CT-2000-00148.
O.V. acknowledges partial support from the Spanish MCYT FPA2002-00612 and 
DGEUI of the Generalitat Valenciana grant GV01-94. O.V. thanks V. Gimenez,
A. Pich and A. Santamaria for useful discussions.  L. V-S. thanks 
CONACyT-Mexico for funding through the scholarship 120436/122667 
and N.T. Leonardo for useful discussions.

\appendix

\section{Experimental Constraints\label{sec:expconst}}

\begin{figure*}[ht]
\vspace*{-0.4cm}
\begin{center}
\mbox{\epsfig{file=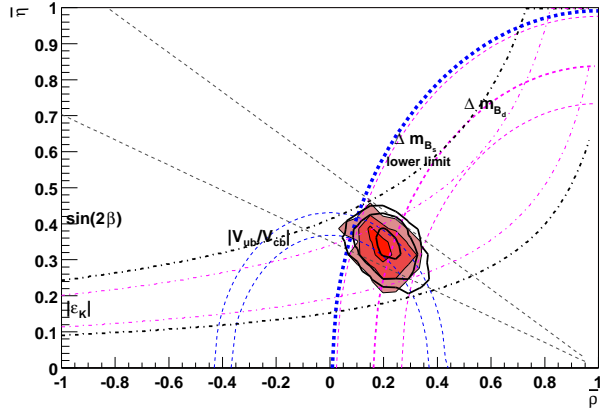,height=6cm}}
\end{center}
\vspace*{-0.75cm}
\caption{\footnotesize{Fits of the parameters $\rhobar$, $\etabar$ to the 
2003 values of the constraints afecting the CKM matrix in the SM and in 
the MSSM case. The dotted lines indicate the regions 
of 1$\sigma$ for the different constriants, except in the case of  
$\sin 2\beta$ where we have ploted the 2$\sigma$ region as a reference.
In $|\epsilon_K|$ the red(light) lines represent this constraint in the SM 
without new physics contributions. The grey(dark) lines represent the 
constraint in the MSSM case where we enlarge the error a 20\% 
as explained in the text.
The coloured (different grey tone) regions are the CL at 99\%, 95\% and 
68\% in the SM fit, while 
the wide continuous lines are the same CL in the MSSM case.}}
\label{fig:fcls2003s20pA}
\end{figure*}

\begin{figure*}[ht]
\vspace*{-0.4cm}
\begin{center}
\mbox{\epsfig{file=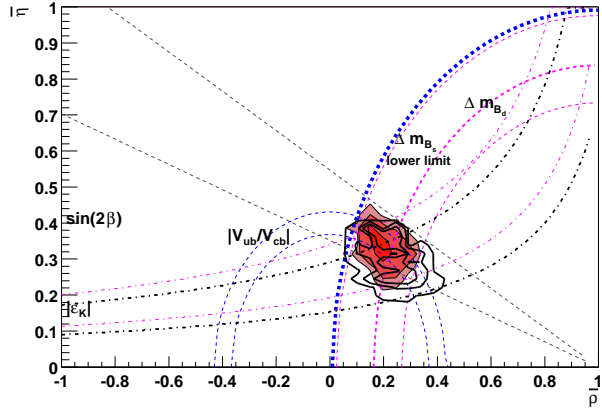,height=6cm}}
\end{center}
\vspace*{-0.75cm}
\caption{\footnotesize{In this figure we assume that $20\%$ of the measured 
value of $\epsilon_k$ is due to SUSY contributions and we fit only the 
remaining $80\%$ of $\epsilon_k$. 
As we can see in this case we can reach lower values of $\etabar$, required 
for lower values of $\Phi_2$.}}
\label{fig:fcls2003s20pC}
\end{figure*}

We have performed a fit to the CKM elements ($A$, $\lambda$, $\bar
\rho$ and $\bar \eta$) with the experimental information available in
Summer 2003, as appear in Table \ref{tab:experiment}. In Table 1 of
\cite{romanino} you can find other parameters, entering the formulas
in the fit, which are well measured and we take as fixed in the fit. 
The relevant formulas for this fit can also be found in reference
\cite{romanino}. 
\begin{table*}[h] 
\begin{tabular*}{1.0\textwidth}{@{\extracolsep{\fill}}|c|c|c|c|}
\hline
\multicolumn{4}{|c|}{{\ Fitted Parameters 2003}}\\ \hline 
Parameter & Value & Gaussian-Flat errors. &Referen.\\
$A$ & $0.834\pm 0.036$ & & * \\
$\lambda$ & $0.2240\pm 0.0036$ & & \cite{Battaglia:2003in}\\
$|V_{ub}|$ &$37.35\times 10^{-4}$ &$(\pm 1.69 \pm 5.0)\times
10^{-4}$ & \\
$|V_{cb}|$ &$41.1\times 10^{-3}$& $(1.63\pm 0.8)\times 10^{-3}$ & \\
$B_K$ & $0.86$ &$0.06\pm 0.14$ &\cite{Battaglia:2003in}\\
$m_c^{\star}$ & $(1.3\pm 0.1) ~{\rm GeV}$ & &*\\
$m_t^{\star}$ & $(167\pm 5)~{\rm GeV}$ & & *\\
$\eta_1^{\star}$ & $1.38\pm 0.53 $ & &\cite{Battaglia:2003in}\\
$\eta_3^{\star}$ & $0.47\pm 0.04$ & &\cite{Battaglia:2003in}\\
$\Delta m_{B_d}$ & $(0.502\pm 0.006)~{\rm ps}^{-1}$ &
&\cite{hfagsummer:03} \\
$f_{B_d}\sqrt{B_{B_d}}$ & $0.228~{\rm GeV}$ & $ (0.030\pm 
0.020)~{\rm GeV}$ &$\dagger$ \cite{Battaglia:2003in} \\
$\xi$& $ 1.21 $ &$0.04\pm 0.05$\ & \cite{Battaglia:2003in}\\
$\Delta m_{B_s}>$ & $14.4~{\rm ps}^{-1}$ at 95\% C.L.&
& \cite{hfagsummer:03}\\
\hline
\end{tabular*}
\caption{\footnotesize {Fitted Parameters 2003.  The parameters marked with  
* have been computed as in \cite{romanino} with the new data 
from~\protect\cite{PDG}.  In $f_{B_d}\sqrt{B_{B_d}}$ ($\dagger$) we have 
conservatively doubled the theoretical (flat) error from 
\cite{Battaglia:2003in} to accomodate 
possible large results from unquenched lattice QCD calculations 
\cite{unquenched}}}
\label{tab:experiment}
\end{table*}
Figures \ref{fig:fcls2003s20pA} and \ref{fig:fcls2003s20pC} show the
results of the fits in the ($\rhobar,\etabar$) plane. In all the plots,
we have used the constraints from $|V_{ub}/V_{cb}|$, $\Delta m_{B_s}$ 
(lower limit), $\Delta m_{B_d}$ and $|\epsilon_K|$. The constraint 
from $\sin 2 \beta$ has {\it not} been included in the fit. 

In Figure \ref{fig:fcls2003s20pA}, due to the possible supersymmetric 
contributions to $\epsilon_K$ as presented in Section \ref{sec:SUSY}, 
we have enlarged the experimental error of $\epsilon_K$ a 20 \%  
of its mean value, $\bar \epsilon_K$. Then, the SM contribution lies 
within the range $\bar \epsilon_K \pm 0.20 \bar \epsilon_K$. We have 
considered that $\epsilon_K$ has a uniform probability within this range. 
However, we do not find a significant variation of $\rhobar$ and $\etabar$, 
with respect to the original SM fit. Hence the phase $\Phi_2$ does not 
change considerably in this case. 

In Figure \ref{fig:fcls2003s20pC}, we have assumed a fixed sign of the SUSY 
contributions, such that the mean of the 
experimental $\epsilon_K$ receives a SUSY contribution of $20\%$ of its 
value, thus decreasing the SM contribution to $80\%$ of $\bar \epsilon_K$.
In this case, the fitted value of $\bar \eta$ decreases $(\bar\eta = 
0.306^{+0.050}_{-0.052})$ with respect to the value $(\bar\eta = 
0.328^{+0.037}_{-0.036})$ obtained in the original SM fit, allowing for 
the lower values of $\Phi_2$ ($=0.599^{+0.350}_{-0.334}$). 

\section{Neutrino mass matrix, Model I \label{sec:neutmamix}}
With the seesaw mechanism, we obtain the low energy effective
neutrino mass matrix $\chi^\nu$ as,
\beq 
\chi^\nu =Y^\nu M^{-1}_R
(Y^\nu)^T = Y^\nu L_M (M_R)^{-1}_{\rm diag} L^T_M (Y^\nu)^T, 
\eeq 
where $L_M$ is the complex orthogonal matrix diagonalising
$M_R$ and up to ${\cal{O}}(\varepsilon^{5.5})$,
\bea
L_M &\simeq&\left(\begin{array}{ccc}i c_{12}e^{-i(\delta^\prime - 4 \alpha_u)}&
s_{12} e^{-i(\delta^\prime - 4 \alpha_u)}&0\\ -i s_{12} e^{-i 4 \alpha_u}  &
c_{12} e^{-i 4 \alpha_u}  &\eta_1
\varepsilon^{2.75} e^{-i 4 \alpha_u}\\ i s_{12}\eta_1 
\varepsilon^{2.75}& -c_{12}\eta_1\varepsilon^{2.75} &
1\end{array} \right)
\eea
where $s_{12} \simeq \frac{\lambda}{\rho} \varepsilon^{1/4}$ and 
$c_{12} \simeq 1 - \frac{\lambda^2}{2\rho^2} \varepsilon^{1/2}$. Now
$(M_R)^{-1}_{\rm diag}= \mbox{Diag.}(\frac{\rho}{\lambda^2 \varepsilon^6},
\frac{1}{ \rho \varepsilon^{5.5}}, 1)$ and
\bea
\label{dirac'}
(Y^{\nu})^\prime~~= ~~~~ Y^\nu \cdot L_M~~ \simeq ~~ 
\left(\begin{array}{ccc}e^{i(\delta^u - 4 \alpha_u)}&&\\&e^{-i \beta_3}&\\&&1
\end{array} \right)~ \cdot~~~~
~~~~~~~~~~~~~~~~~~~~~~~~~\nn \\
\left(
\barr{ccc}
- i s_{12} \varepsilon^3 (X_1 - X_2 \eta_1
\varepsilon^{2.75} e^{i \varphi}) &  c_{12} \varepsilon^3 (X_1 - X_2 \eta_1
\varepsilon^{2.75} e^{i \varphi}) & \varepsilon^3 
(X_2 e^{i\varphi} + X_1 \eta_1 \varepsilon^{2.75})\\
i \varepsilon^3 (c_{12}X_1 + s_{12} \eta \eta_1
\varepsilon^{2.5} e^{i \varphi}) & \varepsilon^3 (s_{12} X_1 - c_{12} \eta 
\eta_1\varepsilon^{2.5} e^{i \varphi}) & \eta \varepsilon^{2.75} e^{i\varphi}\\
i c_{12} \varepsilon^3 X_2 - i s_{12} (\eta -\eta_1)
\varepsilon^{2.75} & s_{12} \varepsilon^3 X_2 + c_{12}(\eta -  
\eta_1) \varepsilon^{2.75} & 1 \\
\earr
\right),
\eea
where we have defined $\rho = (\eta_2 - \eta_1^2)$,
$X_1 = (1 +X)$, $X_2 = (1-X)$ and the phase $\varphi = 4 \alpha_u + \beta_3$.
Notice that the structure of the rotated Yukawa matrix has allowed 
us to extract a diagonal matrix of phases such that only $\varphi$ appears
in the remaining matrix. 
Now we can obtain the elements of the effective neutrino
mass matrix $m^\nu_{LL} =  v_2^2 P\cdot \chi^\nu \cdot P$ with $P =
\mbox{Diag}(e^{i(\delta^u - 4 \alpha_u)},e^{-i \beta_3},1)$ and,
\bea
\label{chiinterm}
&\chi^\nu \simeq & \\[.2cm]
& \left(\begin{array}{ccc}- \frac{\rho}{\lambda^2} s_{12}^4 X_1 &\frac{\rho}{\lambda^2} 
s_{12} c_{12} X_1^2 (1+s_{12}^2) & \frac{\rho}{\lambda^2} s_{12} X_1 \left(
c_{12} X_2 (1+s_{12}^2) - \frac{(\eta- \eta_1) \lambda}{\rho} 
s_{12}^2 \right) \\ & - \frac{\rho}{\lambda^2} X_1^2 c_{12}^2
&- \frac{\rho}{\lambda^2} \left(X_1 X_2 c_{12}^2
- \frac{(\eta- \eta_1) \lambda}{\rho} c_{12} (1 + s_{12}^2) X_1
\right) \\ & & 1  + 2 \frac{(\eta-\eta_1 )}{ \lambda}c_{12} 
(1 + s_{12}^2) X_2  - \frac{\rho}{\lambda^2} c_{12}^2
X_2^2 - \frac{(\eta-\eta_1 )^2}{\rho} s_{12}^2\end{array} \right)& \nn
\eea

However this is the neutrino mass matrix before diagonalisation of
charged lepton Yukawas.
To obtain the lepton mixings and especially the $CP$ violating phases 
relevant for phenomenology, it is more
convenient to go to the basis of real and diagonal charged lepton 
masses. To reach this basis we must rotate and rephase this effective
neutrino mass matrix: $\chi^\prime = L^{l\dagger}\cdot P \cdot 
\chi\cdot P \cdot L^{l*}$, with,
\bea
\label{chargedrotation}
L^l \equiv P^\prime\cdot U =
\left(\begin{array}{ccc}e^{i\delta^d}&&\\&1&\\&&e^{i(2 \chi - \beta_3)}\end{array} \right)
\cdot \left(\begin{array}{ccc}1  &\bar \varepsilon
\frac{X_1}{\Sigma_e}& 0 \\ -\bar \varepsilon \frac{X_1}{\Sigma_e}&
1 & 0 \\ 0 & 0 &1\end{array} \right) + {\cal{O}} (\bar \varepsilon^2)
\eea
which is enough for our purposes. So, we have,
\bea
\label{chifin}
\chi^\prime = \tilde P  \cdot \left(\begin{array}{ccc}1  &- \bar \varepsilon
\frac{X_1}{\Sigma_e} e^{i \varphi_L}
& 0 \\ \bar \varepsilon \frac{X_1}{\Sigma_e}
e^{-i \varphi_L}&
1 & 0 \\ 0 & 0 &1\end{array} \right) \cdot \chi \cdot \left(\begin{array}{ccc}1  &\bar \varepsilon
\frac{X_1}{\Sigma_e} e^{-i\varphi_L} & 0 \\ 
-\bar \varepsilon \frac{X_1}{\Sigma_e}e^{i \varphi_L}&
1 & 0 \\ 0 & 0 &1\end{array} \right) \cdot \tilde P 
\eea
where $\varphi_L = -2 \alpha_d - 2 \alpha_u+ \beta_3$ and 
$\tilde P = \mbox{Diag}(e^{i(-2 \alpha_d -2 \alpha_u)},
e^{-i \beta_3},e^{i(-2 \chi + \beta_3)})$


\begin{thebibliography}{99}

\bibitem{graham}
For a review and further references see:\\
G.~G.~Ross, ``Models of Fermion masses'', Published
in \textit{TASI 2000} ed. J.~L.~Rosner (World Scientific,
New Jersey, 2001)

\bibitem{annrev}
A.~Masiero and O.~Vives,
Ann.\ Rev.\ Nucl.\ Part.\ Sci.\  {\bf 51} (2001) 161
[arXiv:hep-ph/0104027].


\bibitem{king}
S.~F.~King and G.~G.~Ross,
Phys.\ Lett.\ B {\bf 520} (2001) 243
[arXiv:hep-ph/0108112].


\bibitem{Ross:2002fb}
G.~G.~Ross and L.~Velasco-Sevilla,
Nucl.\ Phys.\ B {\bf 653} (2003) 3
[arXiv:hep-ph/0208218].

\bibitem{king2}
S.~F.~King and G.~G.~Ross,
arXiv:hep-ph/0307190.


\bibitem{romanino}
R.~G.~Roberts, A.~Romanino, G.~G.~Ross and L.~Velasco-Sevilla,
Nucl.\ Phys.\ B {\bf 615} (2001) 358
[arXiv:hep-ph/0104088].

\bibitem{dine}
M.~Dine, R.~G.~Leigh and D.~A.~MacIntire,
Phys.\ Rev.\ Lett.\  {\bf 69} (1992) 2030
[arXiv:hep-th/9205011].


\bibitem{spontCPflavour}
Y.~Nir and R.~Rattazzi,
Phys.\ Lett.\ B {\bf 382} (1996) 363
[arXiv:hep-ph/9603233];
\\
J.~L.~Chkareuli, C.~D.~Froggatt and H.~B.~Nielsen,
Nucl.\ Phys.\ B {\bf 626} (2002) 307
[arXiv:hep-ph/0109156].

\bibitem{RRR}
P.~Ramond, R.~G.~Roberts and G.~G.~Ross,
Nucl.\ Phys.\ B {\bf 406} (1993) 19
[arXiv:hep-ph/9303320].

\bibitem{gatto}
R.~Gatto, G.~Sartori and M.~Tonin,
Phys.\ Lett.\ B {\bf 28} (1968) 128.

\bibitem{FN}
C.~D.~Froggatt and H.~B.~Nielsen,
Nucl.\ Phys.\ B {\bf 147} (1979) 277.

\bibitem{softflavour}
M.~Dine, R.~G.~Leigh and A.~Kagan,
Phys.\ Rev.\ D {\bf 48} (1993) 4269
[arXiv:hep-ph/9304299];
\\
A.~Pomarol and D.~Tommasini,
Nucl.\ Phys.\ B {\bf 466} (1996) 3
[arXiv:hep-ph/9507462];
\\
L.~J.~Hall and H.~Murayama,
Phys.\ Rev.\ Lett.\  {\bf 75} (1995) 3985
[arXiv:hep-ph/9508296];
\\
R.~Barbieri, G.~R.~Dvali and L.~J.~Hall,
Phys.\ Lett.\ B {\bf 377} (1996) 76
[arXiv:hep-ph/9512388];
\\
Z.~Berezhiani,
Phys.\ Lett.\ B {\bf 417} (1998) 287
[arXiv:hep-ph/9609342];
\\
E.~Dudas, C.~Grojean, S.~Pokorski and C.~A.~Savoy,
Nucl.\ Phys.\ B {\bf 481} (1996) 85
[arXiv:hep-ph/9606383];
\\
R.~Barbieri, L.~J.~Hall, S.~Raby and A.~Romanino,
Nucl.\ Phys.\ B {\bf 493} (1997) 3
[arXiv:hep-ph/9610449].



\bibitem{alignment}
Y.~Nir and N.~Seiberg,
Phys.\ Lett.\ B {\bf 309} (1993) 337
[arXiv:hep-ph/9304307];
\\
M.~Leurer, Y.~Nir and N.~Seiberg,
Nucl.\ Phys.\ B {\bf 420} (1994) 468
[arXiv:hep-ph/9310320];
\\
Y.~Nir and G.~Raz,
Phys.\ Rev.\ D {\bf 66} (2002) 035007
[arXiv:hep-ph/0206064].

\bibitem{peddie}
E.~Dudas, S.~Pokorski and C.~A.~Savoy,
Phys.\ Lett.\ B {\bf 356} (1995) 45
[arXiv:hep-ph/9504292];
\\
E.~Dudas, S.~Pokorski and C.~A.~Savoy,
Phys.\ Lett.\ B {\bf 369} (1996) 255
[arXiv:hep-ph/9509410];
\\
I.~Jack, D.~R.~T.~Jones and R.~Wild,
Phys.\ Lett.\ B {\bf 580} (2004) 72
[arXiv:hep-ph/0309165];
\\
S.~F.~King and I.~N.~R.~Peddie,
arXiv:hep-ph/0312237.

\bibitem{glosi} S.~F.~King, I.~N.~R. Peddie, G.~G.~Ross, L. Velasco-Sevilla
and O. Vives, work in progress.


\bibitem{leutwyler}
H.~Leutwyler,
Nucl.\ Phys.\ Proc.\ Suppl.\  {\bf 94} (2001) 108
[arXiv:hep-ph/0011049].

\bibitem{toni}
E.~Gamiz, M.~Jamin, A.~Pich, J.~Prades and F.~Schwab,
JHEP {\bf 0301} (2003) 060
[arXiv:hep-ph/0212230].

\bibitem{charm}
 J.~Penarrocha and K.~Schilcher,
Phys.\ Lett.\ B {\bf 515} (2001) 291
[arXiv:hep-ph/0105222];
\\
J.~Erler and M.~x.~Luo,
Phys.\ Lett.\ B {\bf 558} (2003) 125
[arXiv:hep-ph/0207114];
\\
M.~Eidemuller,
Phys.\ Rev.\ D {\bf 67} (2003) 113002
[arXiv:hep-ph/0207237];
\\
B.~L.~Ioffe and K.~N.~Zyablyuk,
Eur.\ Phys.\ J.\ C {\bf 27} (2003) 229
[arXiv:hep-ph/0207183].

\bibitem{arcadi}
G.~Rodrigo, A.~Pich and A.~Santamaria,
Phys.\ Lett.\ B {\bf 424} (1998) 367
[arXiv:hep-ph/9707474].

\bibitem{MI}
F.~Gabbiani, E.~Gabrielli, A.~Masiero and L.~Silvestrini,
Nucl.\ Phys.\ B {\bf 477} (1996) 321
[arXiv:hep-ph/9604387].

\bibitem{PDG}
K.~Hagiwara {\it et al.}  [Particle Data Group Collaboration],
Phys.\ Rev.\ D {\bf 66} (2002) 010001.

\bibitem{Olechowski:1990bh}\
M.~Olechowski and S.~Pokorski,\
Phys.\ Lett.\ {\bf 257\ } (1991) 388.\

\bibitem{barger}
V.~D.~Barger, M.~S.~Berger and P.~Ohmann,
Phys.\ Rev.\ D {\bf 47} (1993) 2038
[arXiv:hep-ph/9210260].



\bibitem{peccei}
R.~D.~Peccei and H.~R.~Quinn,
Phys.\ Rev.\ Lett.\  {\bf 38} (1977) 1440.

\bibitem{Velasco-Sevilla:2003gd}
L.~Velasco-Sevilla,
JHEP {\bf 0310} (2003) 035
[arXiv:hep-ph/0307071].

\bibitem{King:2002nf} S.~F.~King, 
Phys.\ Lett.\ B {\bf 439} (1998) 350 [arXiv:hep-ph/9806440]; 
S.~F.~King, 
Nucl.\ Phys.\ B {\bf 576} (2000) 85 [arXiv:hep-ph/9912492]; 
S.~F.~King, 
JHEP {\bf 0209} (2002) 011 [arXiv:hep-ph/0204360]. 

\bibitem{King:2002qh}
S.~F.~King,
Phys.\ Rev.\ D {\bf 67} (2003) 113010
[arXiv:hep-ph/0211228].

\bibitem{sacha}
S.~Davidson and A.~Ibarra,
Nucl.\ Phys.\ B {\bf 648} (2003) 345
[arXiv:hep-ph/0206304];
\\
J.~R.~Ellis and M.~Raidal,
Nucl.\ Phys.\ B {\bf 643} (2002) 229
[arXiv:hep-ph/0206174];
\\
S.~Pascoli, S.~T.~Petcov and W.~Rodejohann,
Phys.\ Rev.\ D {\bf 68} (2003) 093007
[arXiv:hep-ph/0302054];
\\
A.~Ibarra and G.~G.~Ross,
Phys.\ Lett.\ B {\bf 575} (2003) 279
[arXiv:hep-ph/0307051];
\\
S.~Davidson and R.~Kitano,
arXiv:hep-ph/0312007.


\bibitem{tatsuo}
T.~Kobayashi and O.~Vives,
Phys.\ Lett.\ B {\bf 506} (2001) 323
[arXiv:hep-ph/0011200].

\bibitem{khalil}
S.~Abel, S.~Khalil and O.~Lebedev,
Phys.\ Rev.\ Lett.\  {\bf 89} (2002) 121601
[arXiv:hep-ph/0112260].

\bibitem{flat}
G.~G.~Ross and O.~Vives,
Phys.\ Rev.\ D {\bf 67} (2003) 095013
[arXiv:hep-ph/0211279].





\bibitem{barr}
S.~M.~Barr and A.~Masiero,
Phys.\ Rev.\ D {\bf 38} (1988) 366.



\bibitem{piai}
A.~Masiero, M.~Piai and O.~Vives,
Phys.\ Rev.\ D {\bf 64} (2001) 055008
[arXiv:hep-ph/0012096].

\bibitem{oleg}
O.~Lebedev,
Phys.\ Rev.\ D {\bf 67} (2003) 015013
[arXiv:hep-ph/0209023].

\bibitem{reyes}
D.~Becirevic {\it et al.},
Nucl.\ Phys.\ B {\bf 634} (2002) 105
[arXiv:hep-ph/0112303].

\bibitem{phiks}
K.~Abe  [Belle Collaboration],
arXiv:hep-ex/0308035,
\\
B.~Aubert {\it et al.}  [BABAR Collaboration],
arXiv:hep-ex/0207070.

\bibitem{bphiks}
E.~Lunghi and D.~Wyler,
Phys.\ Lett.\ B {\bf 521} (2001) 320
[arXiv:hep-ph/0109149];
\\
T.~Goto, Y.~Okada, Y.~Shimizu, T.~Shindou and M.~Tanaka,
Phys.\ Rev.\ D {\bf 66}, 035009 (2002)
[arXiv:hep-ph/0204081];
\\
D.~Chang, A.~Masiero and H.~Murayama,
Phys.\ Rev.\ D {\bf 67}, 075013 (2003)
[arXiv:hep-ph/0205111];
\\
S.~Khalil and E.~Kou,
Phys.\ Rev.\ D {\bf 67} (2003) 055009
[arXiv:hep-ph/0212023];
\\
S.~Baek,
Phys.\ Rev.\ D {\bf 67}, 096004 (2003)
[arXiv:hep-ph/0301269];
\\
J.~Hisano and Y.~Shimizu,
Phys.\ Lett.\ B {\bf 565}, 183 (2003)
[arXiv:hep-ph/0303071];
\\
K.~Agashe and C.~D.~Carone,
Phys.\ Rev.\ D {\bf 68}, 035017 (2003)
[arXiv:hep-ph/0304229];
\\
G.~L.~Kane, P.~Ko, H.~b.~Wang, C.~Kolda, J.~h.~Park and L.~T.~Wang,
Phys.\ Rev.\ Lett.\  {\bf 90} (2003) 141803
[arXiv:hep-ph/0304239].





\bibitem{ciuchini}
M.~Ciuchini, E.~Franco, A.~Masiero and L.~Silvestrini,
Phys.\ Rev.\ D {\bf 67} (2003) 075016
[arXiv:hep-ph/0212397].

\bibitem{LeptMI}
I.~Masina and C.~A.~Savoy,
Nucl.\ Phys.\ B {\bf 661} (2003) 365
[arXiv:hep-ph/0211283];
\\
J.~I.~Illana and M.~Masip,
Phys.\ Rev.\ D {\bf 67} (2003) 035004
[arXiv:hep-ph/0207328];
\\
M.~Ciuchini, A.~Masiero, L.~Silvestrini, S.~K.~Vempati and O.~Vives,
arXiv:hep-ph/0307191.

\bibitem{KKV}
S.~Khalil, T.~Kobayashi and O.~Vives,
Nucl.\ Phys.\ B {\bf 580} (2000) 275
[arXiv:hep-ph/0003086].

\bibitem{KvsB}
A.~Masiero and O.~Vives,
Phys.\ Rev.\ Lett.\  {\bf 86} (2001) 26
[arXiv:hep-ph/0007320];
\\
A.~E.~Faraggi and O.~Vives,
Nucl.\ Phys.\ B {\bf 641} (2002) 93
[arXiv:hep-ph/0203061].



\bibitem{Battaglia:2003in}
M.~Battaglia {\it et al.},
arXiv:hep-ph/0304132.

\bibitem{hfagsummer:03}
Heavy Flavor Averaging Group. Results Summer 2003. 
Available at http://www.slac.stanford.edu/xorg/hfag/

\bibitem{unquenched}
D.~Becirevic,
arXiv:hep-ph/0310072.

\end{thebibliography}
\end{document}